\documentclass[12pt]{article}
\usepackage[utf8]{inputenc}
\usepackage[margin=1in]{geometry}
\usepackage{setspace}
\usepackage{graphicx}
\usepackage{natbib, bbm, bm}
\usepackage{amsmath, amsthm, amssymb}

\usepackage[colorlinks=true, allcolors=blue]{hyperref}

\DeclareMathOperator*{\argmax}{arg\,max}

\title{Simulating flood event sets using extremal principal components}
\author{Christian Rohrbeck$^1$ and Daniel Cooley$^2$}
\date{$^1$ Department of Mathematical Sciences, University of Bath, UK\\
$^2$ Department of Statistics, Colorado State University, CO, USA
}

\providecommand{\keywords}[1]
{
  \small	
  \textbf{Keywords--} #1
}

\begin{document}

\maketitle

\begin{abstract}
Hazard event sets, a collection of synthetic extreme events over a given period, are important for catastrophe modelling. This paper addresses the issue of generating event sets of extreme river flow for northern England and southern Scotland, a region which has been particularly affected by severe flooding over the past 20 years. We start by analysing historical extreme river flow across 45 gauges, located within the study region, using methods from extreme value analysis, including the concept of extremal principal components. Our analysis reveals interesting connections between the extremal dependence structure and the region's topography/climate. We then introduce a framework which is based on modelling the distribution of the extremal principal components in order to generate synthetic events of extreme river flow. The generative framework is dimension-reducing in that it distinctly handles the principal components based on their contribution to describing the nature of extreme river flow across the study region. We also detail a data-driven approach to select the optimal dimension. Synthetic flood events are subsequently generated efficiently by sampling from the fitted distribution. Our approach for generating hazard event sets can be easily implemented by practitioners and our results indicate good agreement between the observed and simulated extreme river flow dynamics. For the considered application, we also find that our approach outperforms existing statistical approaches for generating hazard event sets.
\end{abstract}


\newpage

\section{Introduction}
\label{sec:Intro}

Severe flood events regularly cause widespread disruptions and huge losses. In the UK, the flooding caused by Storm Desmond, Storm Eva and Storm Frank in 2015/2016 led to an estimated economic damage of between £1.3--1.9 billion \citep{UKES2018}, and the cascading effects of Storm Ciara and Storm Dennis in February 2020 broke record levels for multiple rivers. Catastrophe models are an important tool to estimate the impact of such natural hazards \citep{Grossi2005} and are used by insurance companies to predict the financial capital required to cover potential payouts. One component of these models is a set of simulated hazard events, representing, for instance, a collection of potential floods over a long period, e.g., 1,000 or 10,000 years. 

Approaches for generating hazard sets fall into two broad categories: numerical and statistical. Numerical approaches often try to capture the physics of the phenomenon of interest, which requires the modelling of a number of complex processes, such as rainfall and soil conditions. For example, numerical weather models can be used to produce simulated spatio-temporal rainfall, and this can be coupled with runoff models to assess flood hazards \citep[e.g.,][]{Camici2014}.

\begin{figure}[t]
\centering
\includegraphics[width=7.5cm]{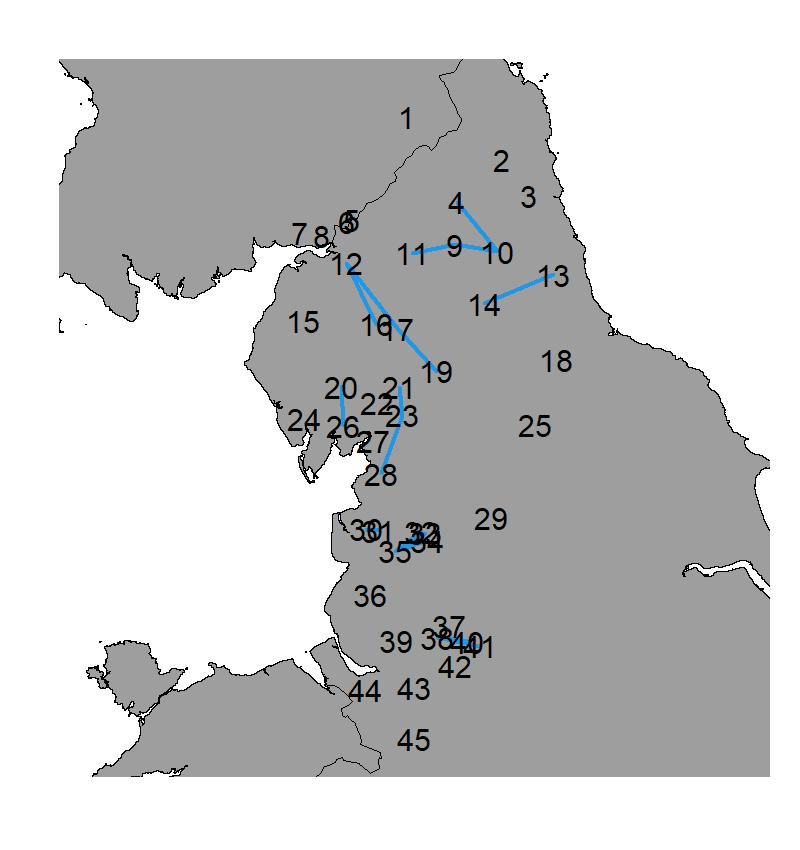}
\hspace{1cm}
\includegraphics[width=7.5cm]{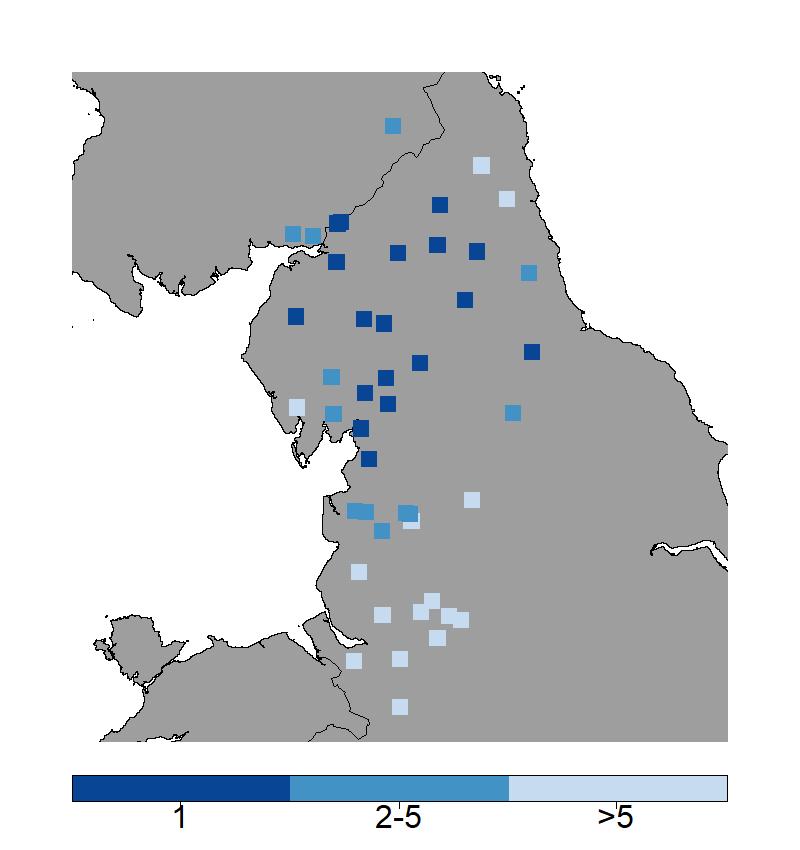}
\caption{Locations of the 45 river flow gauges in northern England and southern Scotland~(left) and the effect of Storm Desmond on river flow levels (right). The numbering in the left plot is determined by the latitude of the gauge, and the blue lines highlight flow-connected gauges. In the right plot, gauges are coloured subject to the rank of the observation recorded during/after Storm Desmond. The darkest coloured gauges recorded their maximum river flow between 1980 and 2018 in the week of Storm Desmond; the second darkest colour corresponds to the observation being amongst the five highest values for the gauge, while the lightest colour indicates that the observed river flow was not very extreme.}
\label{fig:Locations}
\end{figure}

In this paper, we propose a novel statistical framework for hazard set generation in order to produce hazard event sets of extreme river flow for northern England and southern Scotland, one of the UK regions most affected by flooding -- Storm Desmond and Storm Frank led to collapsed bridges and thousands of flooded homes in Cumbria, northern Lancashire and Dumfries and Galloway. To assess the region's historical flood risk, we obtained daily river flow levels (in m$^3$/s) for the period 01/01/1980--30/09/2018 for 45 gauges from the UK's National River Flow Archive (\url{nrfa.ceh.a.uk}). Figure \ref{fig:Locations} left panel shows that most of the gauges are located along the west coast, ranging from southern Scotland to the Welsh border, and in North-East England. The region has a varying topography (described later in Section \ref{sec:Data}) which we will find to be represented by some of our results. To give an example of the spatial structure of extreme river flow, the right panel in Figure~\ref{fig:Locations} highlights the gauges that recorded very extreme flow due to Storm Desmond, most of them are located in Cumbria and northern Lancashire.

Statistical approaches for generating hazard sets often rely on extreme value theory as it provides asymptotically justified methods to analyse the tail behaviour of multivariate random variables \citep{Beirlant2004} and stochastic processes \citep{Davison2012}. Crucially, extremes models provide a framework for extrapolation, i.e., they can estimate the occurrence probability of events outside the range of recorded data. Conceptually, if one can draw realizations from an extremes model, it can be used to generate hazard sets. \citet{Keef2013} and \citet{Quinn2019} produce flood hazard sets using the conditional extremes approach \citep{Heffernan2004}. A potential drawback of their approach is that several residual distributions have to be modelled and the analysis of the tail from the estimated conditional distributions is challenging.

When analysing the tail behaviour of a $K$-dimensional random vector $\mathbf{X}$, marginal distributions and extremal dependence structure are often modelled separately. While a block-maxima or peaks-over threshold model is generally employed for the marginals of $\mathbf{X}$ \citep{Coles2001}, most existing approaches for extremal dependence \citep{Tawn1988, Husler1989, Boldi2007,Cooley2010, Ballani2011, Carvalho2014} are limited to fairly moderate dimensions; see \citep{Engelke2020b} for a review. This limitation is caused by extremal dependence being defined via a measure $H_X$ on the $K$-dimensional unit sphere (or simplex), which usually has to be estimated based on a small number of extreme events. Some issues can be overcome by instead defining a graphical model on the dependence structure \citep{Engelke2020}, but the assumptions on $H_X$ may still be too strong. The extremal dependence in our application is likely to be very complex due to the $K=45$ gauges in Figure~\ref{fig:Locations} being spread across a river network with disparate catchments, and the varying topography and climate across the region.

Rather than beginning with f itting a model, our approach for generating hazard sets starts with an extremal principal component analysis (PCA). Non-extreme PCA is often used as a dimension reducing exploratory tool for high-dimensional data. Recently, both \cite{Cooley2019} and \cite{Drees2021} have adapted ideas from PCA for studying multivariate extremes. Non-extreme PCA has also been applied to define generative models. \cite{Dreveton2004} uses PCA to generate synthetic temperature data. Unlike precipitation and river flow, temperature is approximately Gaussian and well-suited to the elliptical nature of traditional PCA. No analogous procedure exists for generating extreme events; we will later see that the extreme case is more challenging, since the principal components are dependent.

This paper makes two substantial contributions to the area of statistical flood risk analysis. Our first contribution is the analysis of historical extreme river flow in northern England and southern Scotland using recently developed methodology in extreme value analysis, in particular, clustering and PCA. The second contribution is our method to generate hazard event sets. Our framework utilises the methodology by \citet{Cooley2019}, which provides a transformation of $\mathbf{X}$ into a $K$-dimensional random variable $\mathbf{V}$, termed the extremal principal components, with the extremes of $\mathbf{V}$ and $\mathbf{X}$ being linked. Critically, in our river flow application, the first components of $\mathbf{V}$ describe the large-scale spatial structure in the extreme river flow, while the remaining components capture local-scale dynamics or residual behavior. This invites the application of dimension-reduction techniques -- our proposed methods model the full extremal dependence structure of the first components of $\mathbf{V}$, and provide a reasonable fit for the remaining components.  The approach presented herein uses the kernel density estimate for spherical data by \citet{Hall1987} to model the extremal dependence; alternatives are discussed at the end of the paper. From the estimated model, large hazard event sets can be generated with very low computational cost. 

The remainder of this paper is organized as follows: Section~\ref{sec:Analysis} models the gauge-wise extreme values, summarizes the approach by \citet{Cooley2019} and applies it to the river flow data; Section~\ref{sec:Generation} introduces our generative framework; the generated flood event sets for northern England and southern Scotland are analysed in Section~\ref{sec:Results}; we conclude with a discussion in Section~\ref{sec:Discussion}.

\section{Analysis of extreme river flow}
\label{sec:Analysis}

\subsection{Data}
\label{sec:Data}

An exploratory data analysis for the $K=45$ gauges in Figure~\ref{fig:Locations} reveals seasonality in both the magnitude and spatial structure of observed extreme river flow levels. An increase in river flow is largely driven by convectional rainfall (e.g., thunderstorms) in summer months, and by frontal rainfall (e.g, extratropical cyclones) in winter months. Since the most severe flood events occur in winter, we focus on generating hazard event sets for November--March, and both the marginal distributions and tail dependence can be assumed to be stationary.

We obtain $T=848$ weeks of recorded winter river flow between January 1980 and September 2018. For each of the $K$ gauges and $T$ weeks, the maximum daily river flow is stored for analysis, yielding $T$ data points per gauge. Some gauges have a small proportion of missing values (2-3\%), and complete records are available for 28 of the 45 gauges. The average river flow ranges from 2.67~m$^3/$s (gauge 31) up to 151.8~m$^3/$s (gauge 12), indicating substantial differences in catchment sizes. The study region is also quite topographically varied: low mountain ranges dominate the northern half, while the southern half is mostly flatland.
Twelve of the gauges, most of them in the north-east, are located on rivers flowing east towards the North Sea, while the remaining gauges record westward flows into the Irish Sea.

\subsection{Modelling the marginal distributions}
\label{sec:Marginals}

We first model the marginal distributions, and we describe the process for modeling the tail dependence in subsequent subsections. Let $X_{t,k}$ denote the random variable representing the maximum river flow for gauge $k$ in week $t$ ($k=1,\ldots,K;~t=1,\ldots,T$). We adopt a peaks-over threshold approach \citep{Pickands1975} to model the upper tail of $X_{t,k}$. For some suitably high threshold $u_k$, exceedances by $X_{t,k}$ of $u_k$ are modelled as generalized Pareto distributed, GPD$(\sigma_k,\xi_k)$, with
\begin{equation}
\mathbb{P}(X_{t,k}>x+u_k\mid X_{t,k}>u_k) = \left(1+\frac{\xi_k x}{\sigma_k}\right)_{+}^{-1/\xi_k}\qquad(x>0),
\label{eq:GPD}
\end{equation}
where $(z)_+=\max\{z,0\}$, and ($\sigma_k,\xi_k)\in \mathbb{R}_+\times\mathbb{R}$ are termed the scale and shape parameters respectively. The value of $\xi_k=0$, interpreted as the limit of \eqref{eq:GPD} as $\xi_k\to0$, gives the exponential distribution, whilst $\xi_k<0$ corresponds to a short-tailed distribution with finite upper end point, and $\xi_k >0$ gives a power-law tail decay. 

We select the threshold $u_k$ using graphical diagnostic tools \citep{Coles2001}; see \citet{Wadsworth2016} and \cite{Northrop2017} for recent reviews. The diagnostic plots suggest setting $u_k$ to the empirical 94\% quantile of $X_{t,k}$, leaving about 50 data points at each location. Initial gauge-wise maximum likelihood estimates for $\xi_k$ have a large range between $-0.3$ and $0.58$, and the standard errors of $\sigma_k$ and $\xi_k$ are also large. It appears that estimates for several sites are highly influenced by Storm Desmond in 2015; \citet{Barlow2020} also find that this single event led to quite different tail estimates.

In order to reduce uncertainty in the parameter estimates, we aim to borrow statistical information across gauges. One widely applied approach in flood risk analysis is to group gauges based on catchment attributes and to assume a common shape parameter for all gauges within a group; see, e.g.~\cite{Flood1975} and \cite{Asadi2015}. However, the $K$~gauges in our application are located across several river systems, leading to the groups derived based on catchment attributes to be small. Alternatively, Bayesian hierarchical models have also been proposed to estimate spatially varying parameters in extreme value analysis \citep{Cooley2007, Bracken2016}. The difficulty with this approach in our application is the selection of the spatial priors, because the variations in climate and topography across the study region may lead to considerable differences in the distribution of extreme river flow, even for spatially close gauges.

We propose a new two-step process which derives estimates for $\sigma_k$ and $\xi_k$ ($k=1,\ldots,K$) under the sole assumption that the pooled groups of gauges are contiguous. In the first step, the Bayesian clustering framework by \citet{RohrbeckTawn2020} is used to estimate the gauge-wise shape parameters. In the second step, the scale parameter $\sigma_k$ is estimated using maximum likelihood estimation, with $\xi_k$ being fixed to its posterior mean estimate obtained in the first step. The estimates for $\xi_k$ have a reduced range between $0.10$ and $0.22$, and $\sigma_k$ is estimated based on the observations exceeding the empirical 96\% quantile.

Under the assumption that $\mathbb{P}(X_{k,t}>u_{k})$ is constant for all $t$, the $\tau$-year event is given by
$
u_k + \frac{\sigma_k}{\xi_k}\left[\, \left(\lambda_{u_k}\tau \right)^{\xi_k} - 1\,  \right],
$
where $\lambda_{u_k}$ is the expected number of times $X_{t,k}$ exceeds $u_k$ per year. 
Regulators can require that structures withstand a specific $\tau$-year event (e.g., 1000-year event).
Here, the $\tau$-year event has to be seen as a measure of severity of events in the near future, and not as a prediction for extreme river flow over the next $\tau$ years, since climate change may affect the distribution of extreme river flow in the coming decades. For our application, we find the stationarity assumption to be reasonable for the November--March period over the years 1980--2018. We will discuss in Section~\ref{sec:Discussion} how our analysis, including the generative procedure in Section~\ref{sec:Generation}, can be modified should the distribution of river flow levels be non-stationary.

\subsection{Analysing extremal dependence using principal components}
\label{sec:Cooley2019}

In Section~\ref{sec:PCAResults}, we will apply the method by \cite{Cooley2019} to analyze extremal dependence across the components of the random vector~$\mathbf{X}_t=(X_{t,1},\ldots,X_{t,K})$ representing river flow on day $t=1,\ldots,T$ across the $K$ gauges. In this section we summarize their methodology, and compare it to the approach by \citet{Drees2021}.

Let $\tilde{\mathbf{X}}=\left(\tilde{X}_1,\ldots,\tilde{X}_K\right)$ be a $K$-dimensional random vector, with the marginal distributions given by $\mathbb{P}\left(\tilde{X}_k\leq x\right) = \exp\left(-x^{-2}\right)$ ($x>0; k=1,\ldots,K$), i.e., $\tilde{X}_k$ follows a Fr{\'e}chet distribution. It is further assumed that $\tilde{\mathbf{X}}$ is regularly varying with index $\alpha=2$, that is, for any Borel set $B\subset \mathbb{S}^{K-1}_+=\{\bm{\omega}\in\mathbb{R}_+^K:||\bm{\omega}||_2=1\}$,
\begin{equation}
\lim_{r\to\infty}\mathbb{P}\left(||\tilde{\mathbf{X}}||_2 > rz, \frac{\tilde{\mathbf{X}}}{||\tilde{\mathbf{X}}||_2} \in \mathcal{B}~ \left|~ ||\tilde{\mathbf{X}}||_2 > r \right. \right) 
= z^{-2} H_X(\{B\}),
\label{eq:RV2}
\end{equation}
where $H_X$ is termed the angular measure; we apply a marginal transformation to the random vector $\mathbf{X}_t~(t=1,\ldots,T)$ of river flows later in Section~\ref{sec:PCAResults} to meet these conditions.

Tail dependence of $\tilde{\mathbf{X}}$ is summarized via the $K\times K$ tail pairwise dependence matrix (TPDM)~$\Sigma$, which is defined by the second-order properties of $H_X$. Formally, the $(i,j)$-th element of $\Sigma$ is
\begin{equation}
\Sigma_{i,j} = \int_{\mathbb{S}^{K-1}_+} \omega_i \omega_j\, \mathrm{d}H_X(\bm{\omega})\qquad(i,j=1,\ldots,K).
\label{eq:TPDM}
\end{equation}
Equation \eqref{eq:TPDM} corresponds to the extremal dependence measure of \cite{Larsson2012}. The restriction to $\alpha=2$ gives $\Sigma$ properties analogous to a covariance matrix: $\Sigma$ is positive semidefinite and high values of $\Sigma_{i,j}$ indicate strong extremal dependence of the variables $\tilde{X}_i$ and $\tilde{X}_j$, while values close to zero represent weak or no extremal dependence between $\tilde{X}_i$ and $\tilde{X}_j$. 

Since $\Sigma$ is positive semidefinite and symmetric, we can derive its eigendecomposition and express it in the form $\Sigma=\mathbf{U} \mathbf{D} \mathbf{U}^{\mathrm{T}}$, where $\mathbf{D}$ is a diagonal matrix with entries $\lambda_1\geq\ldots\geq\lambda_K\geq0$, and $\mathbf{U}$ is a $K\times K$ unitary matrix. Analogue to principal component analysis, extremal dependence of the components of $\tilde{\mathbf{X}}$ is explored by investigating the eigenvalue/eigenvector pairs $(\lambda_1,\mathbf{U}_{\cdot,1}), \ldots, (\lambda_K,\mathbf{U}_{\cdot,K})$ sequentially. The extremal principal components of $\tilde{\mathbf{X}}$ are then defined as
\begin{equation}
\mathbf{V} = \mathbf{U}^{\mathrm{T}} \tau^{-1}\left(\tilde{\mathbf{X}}\right),
\label{eq:V}    
\end{equation}
where $\tau^{-1}(\cdot) = \log\left[\exp(\cdot)-1\right]$ is applied component-wise, and the components of $\mathbf{V}$ can take any real value. Lemma A4 in \citet{Cooley2019} implies that $\mathbf{V}$ is regularly varying with $\alpha=2$ and we denote its angular measure by $H_V$. This result will form the basis for our generative framework in Section~\ref{sec:Generation}. Note that the measure $H_V$ operates on the whole unit sphere $\mathbb{S}^{K-1}=\{\mathbf{w}\in\mathbb{R}^K:||\mathbf{w}||_2=1\}$, unlike $H_X$, which is the restricted to the first quadrant. Furthermore, Proposition 6 in \citet{Cooley2019} implies that the TPDM $\tilde\Sigma$ for $\mathbf{V}$ satisfies $\tilde\Sigma_{i,i} = \lambda_i$ and $\tilde\Sigma_{i,j} = 0$ for $i\neq j$. As such, the extremal principal components have analogous properties to the classical principal components; however, the random variables $V_i$ and $V_j$ are not independent. Consequently, the angular measure $H_V$ has no simple form, and we have to estimate it in order to sample realizations of $\mathbf{V}$. We will address the estimation of $H_V$ later in Section~\ref{sec:Generation}.

\citet{Drees2021} consider a slightly more general framework to \eqref{eq:RV2} and \eqref{eq:TPDM}. In particular, they do not require $\alpha = 2$, and $H_X$ operates on $\mathbb{S}^{K-1}$ and not $\mathbb{S}_+^{K-1}$, i.e., the components of $\tilde{\mathbf{X}}$ can take any real value. Extremal dependence is summarized via the limit of $\mathbf{S}=\mathbb{E}(\tilde{\mathbf{X}}\tilde{\mathbf{X}}^{\mathrm{T}} / ||\tilde{\mathbf{X}}||_2^2)$ as $||\tilde{\mathbf{X}}||\to\infty$, which is equivalent to the definition of the TPDM in~\eqref{eq:TPDM}. Using the eigendecomposition of $\mathbf{S}$, $H_X$ is projected onto a lower-dimensional linear subspace of~$\mathbb{S}^{K-1}$. As such, \citet{Drees2021} explore the dimension reduction aspect of principal component analysis, and they provide 
theoretical results for the approximation error of the projection. Similar theoretical results can be shown for the approach of \citet{Cooley2019} using (i) the theoretical arguments by \citet{Drees2021} and (ii) that the function $\tau$ in \eqref{eq:V} has only a negligible effect on the upper tail of the random vector.

\subsection{Application to the UK river flow data}
\label{sec:PCAResults}

The two variables $\tilde{X}_i$ and $\tilde{X}_j$ are asymptotically independent if $\mathbb{P}(\tilde{X}_j > x \mid \tilde{X}_i > x) \rightarrow 0$ as $x \rightarrow \infty$~($i,j=1,\ldots,K$). The first-order approximation described in \eqref{eq:RV2}, on which our method relies, is useful in the case of asymptotic dependence, where the pairwise relationships described by $H_X$ are non-degenerate. No extremal dimension reduction procedures analogous to \citet{Cooley2019} or \citet{Drees2021} have been suggested for the more nuanced case of asymptotic independence. We examine our data and find an assumption of asymptotic dependence is reasonable. Estimates $\hat\Sigma_{i,j}$ are sufficiently different from zero, and likewise estimates for the extremal coefficient of \citet{Schlather2003} are sufficiently different from two, the values corresponding to asymptotic independence. An assumption of asymptotic independence for a larger study region would be more questionable.

We apply a marginal transformation to the components of $\mathbf{X}_t$ in order to obtain a variable $\tilde{\mathbf{X}}_t=(\tilde{X}_{t,1},\ldots,\tilde{X}_{t,K})~ (t=1\ldots,T)$ with marginal distributions as required in Section~\ref{sec:Cooley2019}. Define
\begin{equation}
\tilde{X}_{t,k} = \left[-\log\hat{F}_k(X_{t,k})\right]^{-1/2}\qquad(k=1,\ldots,K),
\label{eq:Transform}
\end{equation}
where $\hat{F}_{k}$ is an estimate for the cumulative distribution function (cdf) of $X_{t,k}$. Applying the method by \cite{ColesTawn1991}, $\hat{F}_k$ is set to the empirical cdf for values below a threshold $u_k$, and a GPD$(\sigma_k,\xi_k)$ is used to model $\hat{F}_k(\cdot)$ above $u_k$. The values $\sigma_k$ and $\xi_k$ are replaced by their estimates obtained in Section~\ref{sec:Marginals}, and $u_k$ is set to the empirical 96\% quantile. 

Let $\mathcal{X}=\{\tilde{\mathbf{x}}_{t} = (\tilde{x}_{t,1},\ldots,\tilde{x}_{t,K}) : t=1,\ldots,T\}$ be the set of marginally transformed observations obtained by applying \eqref{eq:Transform} to the original weekly river flow measurements. We remove the weeks with missing data from $\mathcal{X}$, i.e., $\tilde{\mathbf{x}}_{t}$ is removed if no river flow is recorded for at least one gauge in week~$t$. This leaves about 92\% of observations for the analysis of the extremal dependence structure, and the set of weeks with complete records is denoted by $\mathcal{T}^*\subset\{1,\ldots,T\}$. Define $r_t = ||\tilde{\mathbf{x}}_t||_2$ and $\omega_{t,k} = \tilde{x}_{t,k}\, /\, r_t$~($t\in\mathcal{T}^*;~k=1,\ldots,K$). The $(i,j)$-th element of the TPDM $\Sigma$, defined in~\eqref{eq:TPDM}, is then estimated as
\begin{equation}
\hat{\Sigma}_{i,j} = \frac{K}{n} \sum_{t\in\mathcal{T}^*} \omega_{t,i}\, \omega_{t,j}\,
\mathbb{I}(r_t>r_0), \qquad(i,j=1,\ldots,K),
\label{eq:Sigmahat}
\end{equation}
where $\mathbb{I}(\cdot)$ denotes the indicator function, and $r_0$ is set to the 94\% quantile of $\{r_t:t\in\mathcal{T}^*\}$, which corresponds to $n=47$ weeks being used to estimate $\Sigma$.

\begin{figure}
\centering
\includegraphics[width=0.425\textwidth]{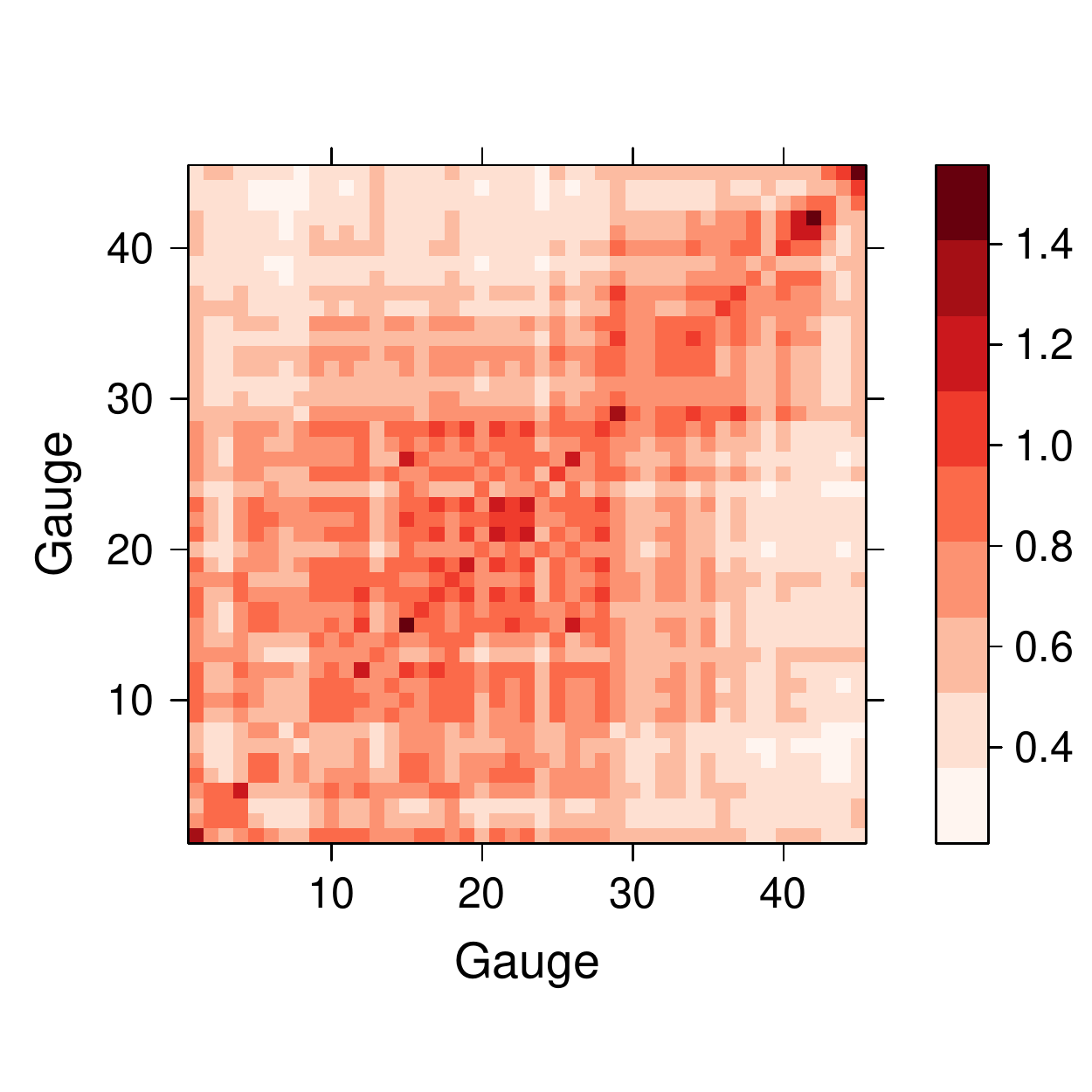}
\hspace{1cm}
\includegraphics[width=0.425\textwidth]{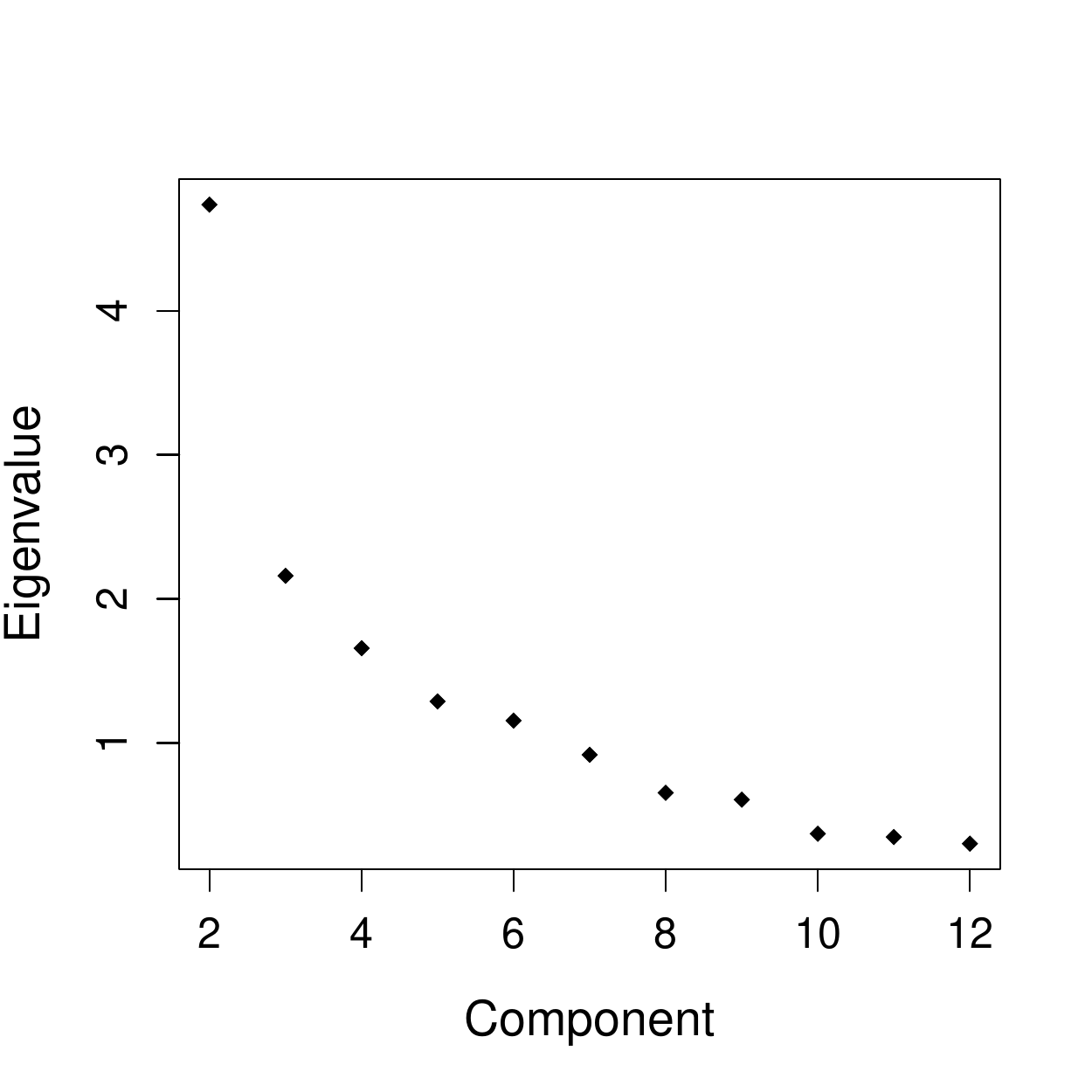}\vspace{-0.5cm}
\caption{Estimated TPDM $\hat\Sigma$ (left) and a scree plot of the eigenvalues $\hat\lambda_2,\ldots,\hat\lambda_{12}$ of $\hat\Sigma$~(right) for the 45 river flow gauges in Figure~\ref{fig:Locations}. The largest eigenvalue $\hat\lambda_1=28.9$ is omitted from the scree plot for visual aid.}
\label{fig:TPDM}
\end{figure}

The estimated TPDM in Figure~\ref{fig:TPDM} left panel indicates that pairwise extremal dependence decreases with increasing spatial distance between gauges; for instance, $\hat\Sigma_{i,j}$ is small when considering pairs of a northern gauge (gauges 1-9) and a southern gauge (gauges 40-45). The block structure apparent between index 28 and 29 corresponds to the separation between the mountainous northern stations and the southern stations.

\begin{figure}
\centering
\includegraphics[width=0.328\textwidth]{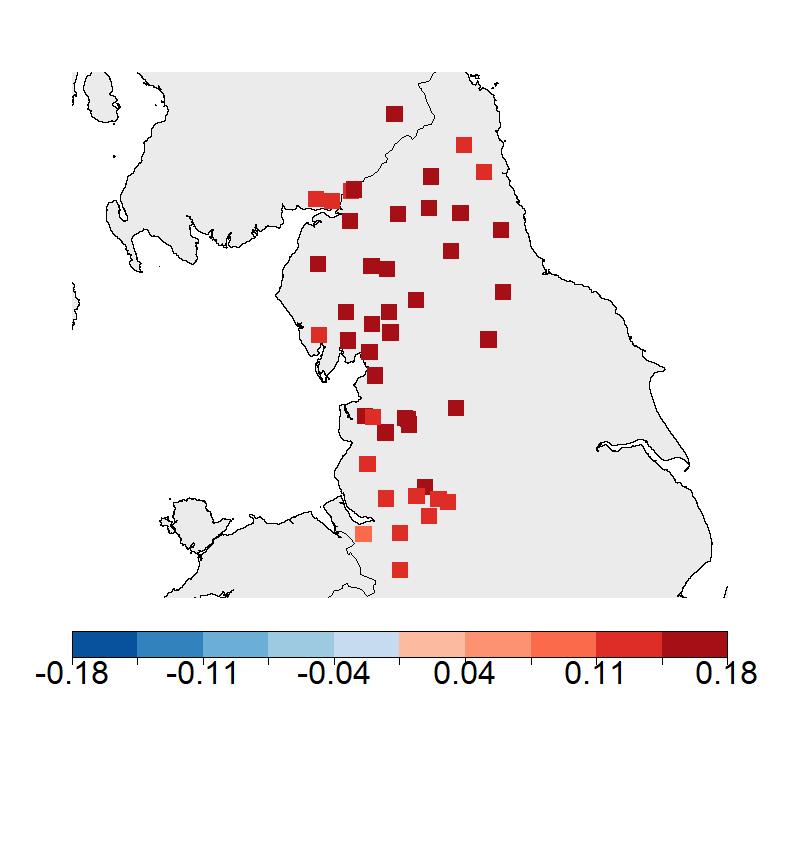}
\includegraphics[width=0.328\textwidth]{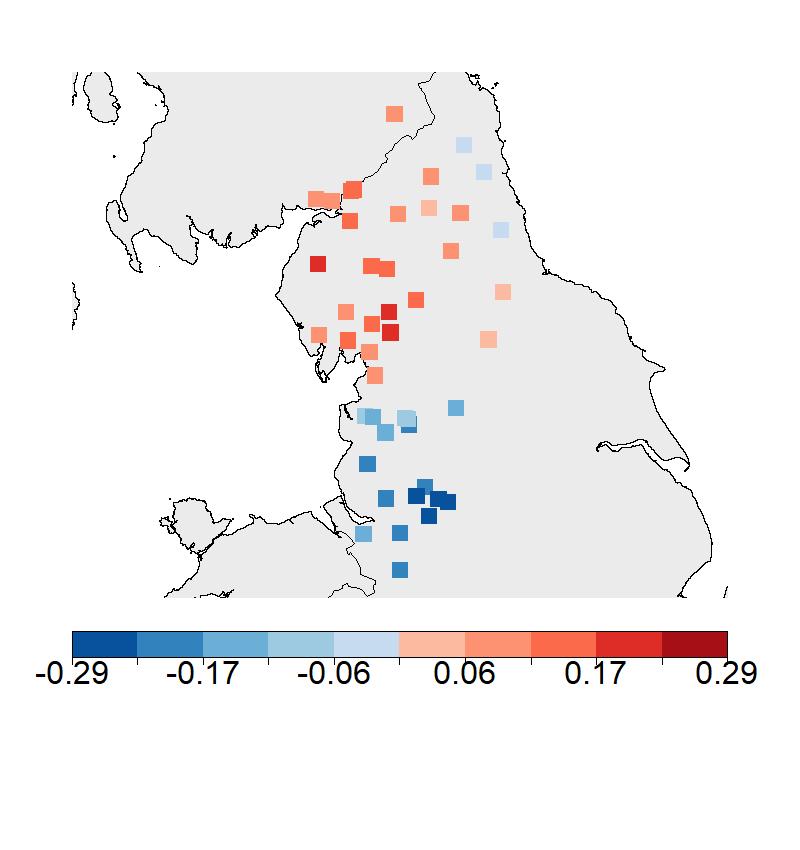}
\includegraphics[width=0.328\textwidth]{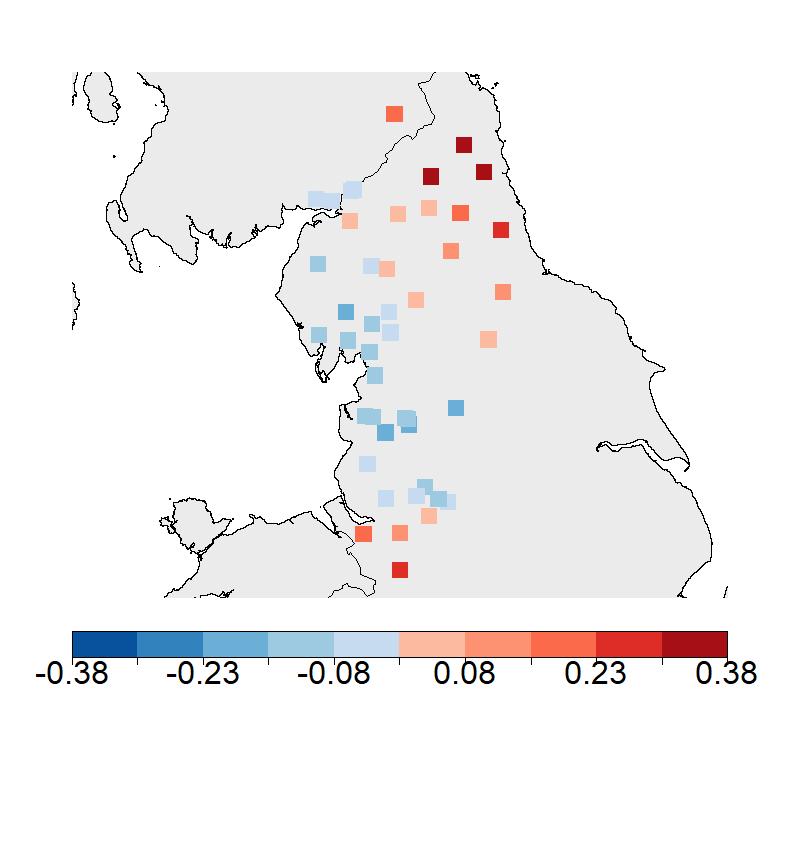}\vspace{-0.8cm}\\
\includegraphics[width=0.328\textwidth]{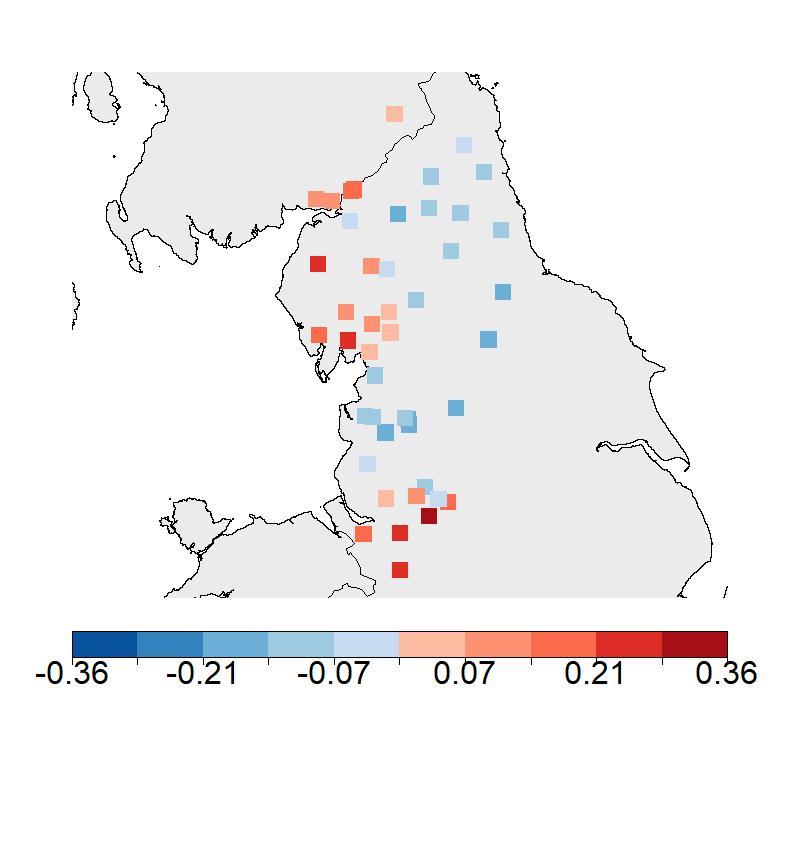}
\includegraphics[width=0.328\textwidth]{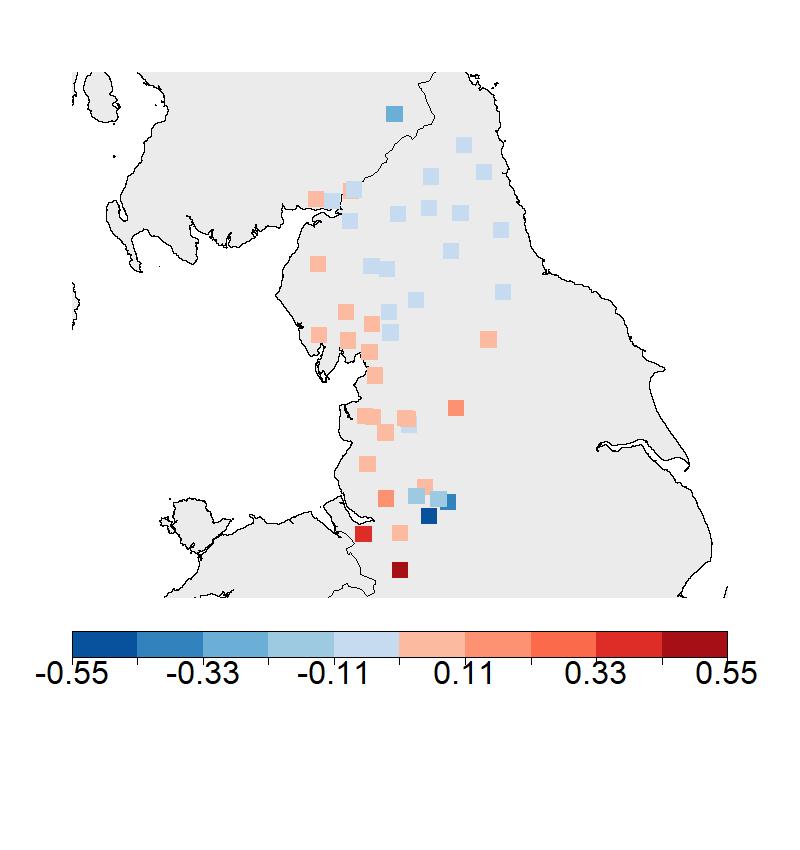}
\includegraphics[width=0.328\textwidth]{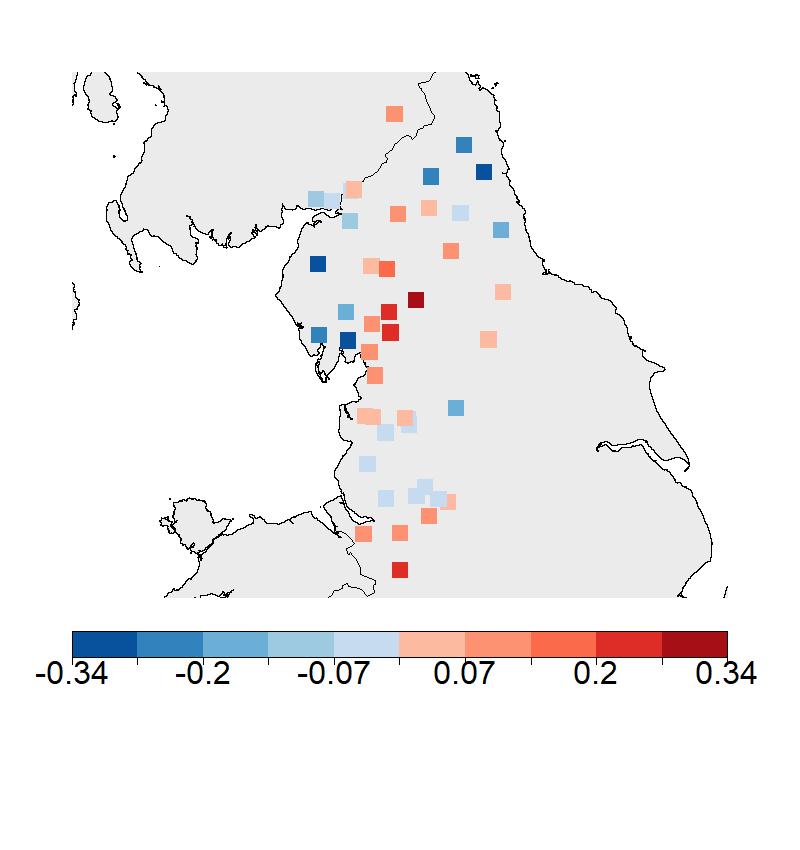}\vspace{-0.8cm}\\
\includegraphics[width=0.328\textwidth]{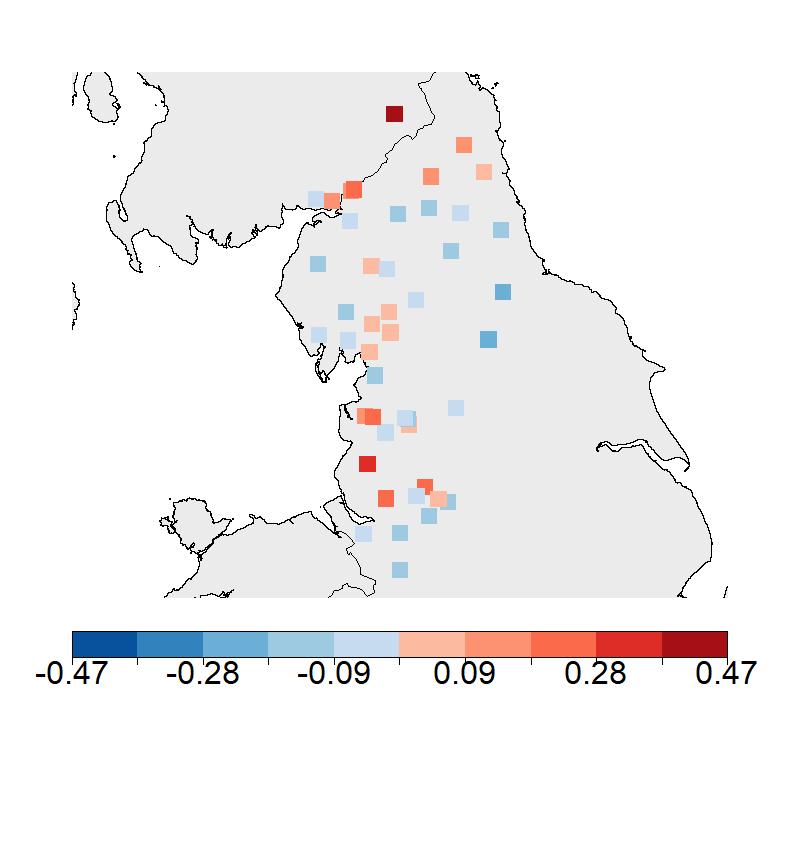}
\includegraphics[width=0.328\textwidth]{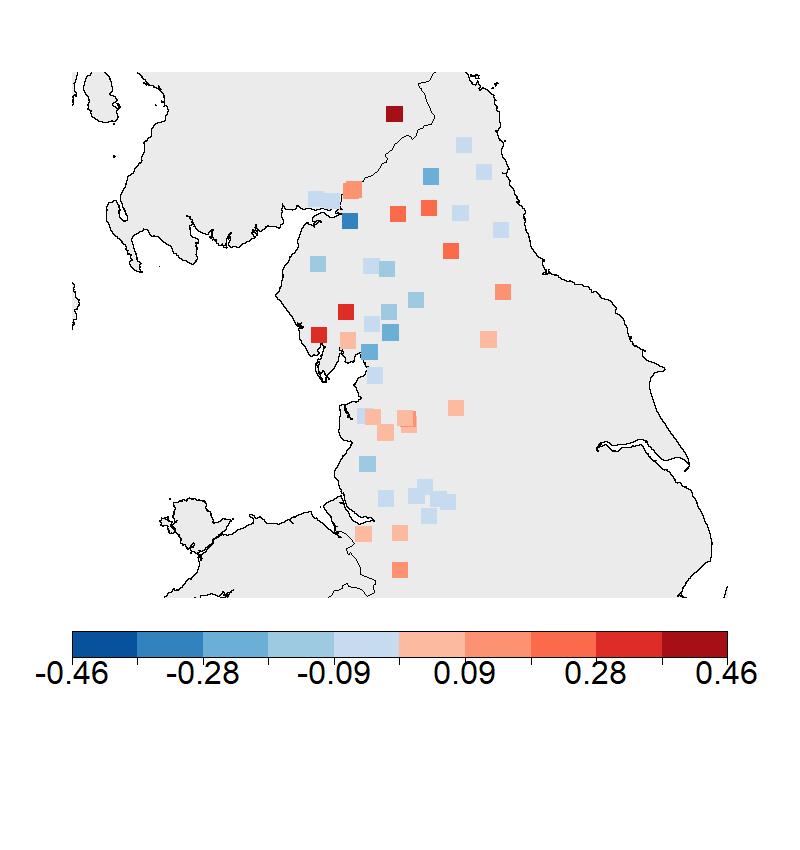}
\includegraphics[width=0.328\textwidth]{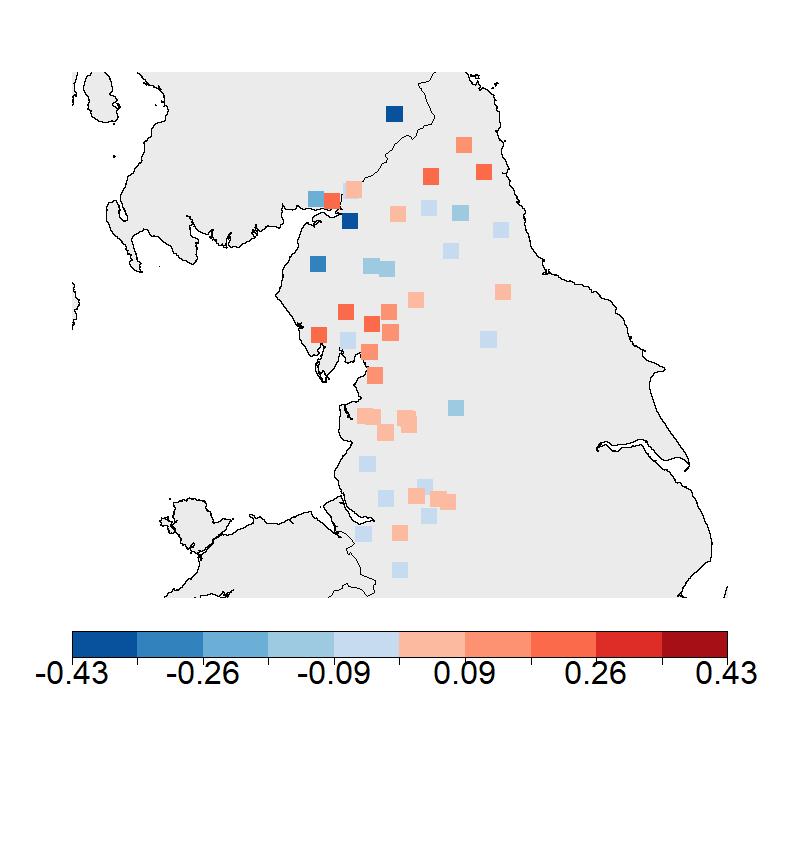}\vspace{-0.8cm}\\
\includegraphics[width=0.328\textwidth]{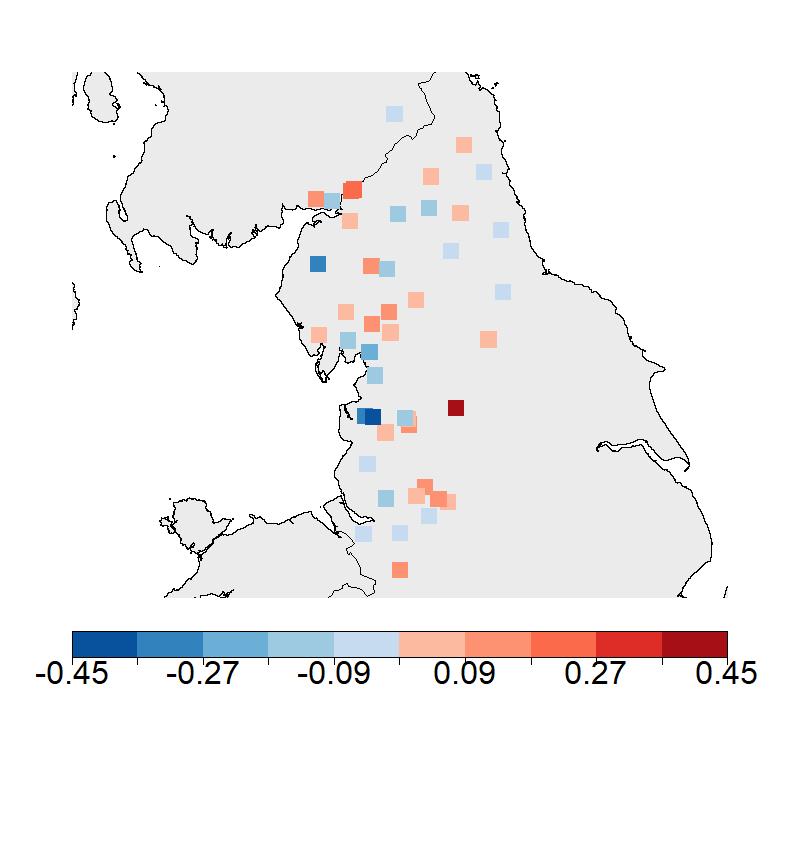}
\includegraphics[width=0.328\textwidth]{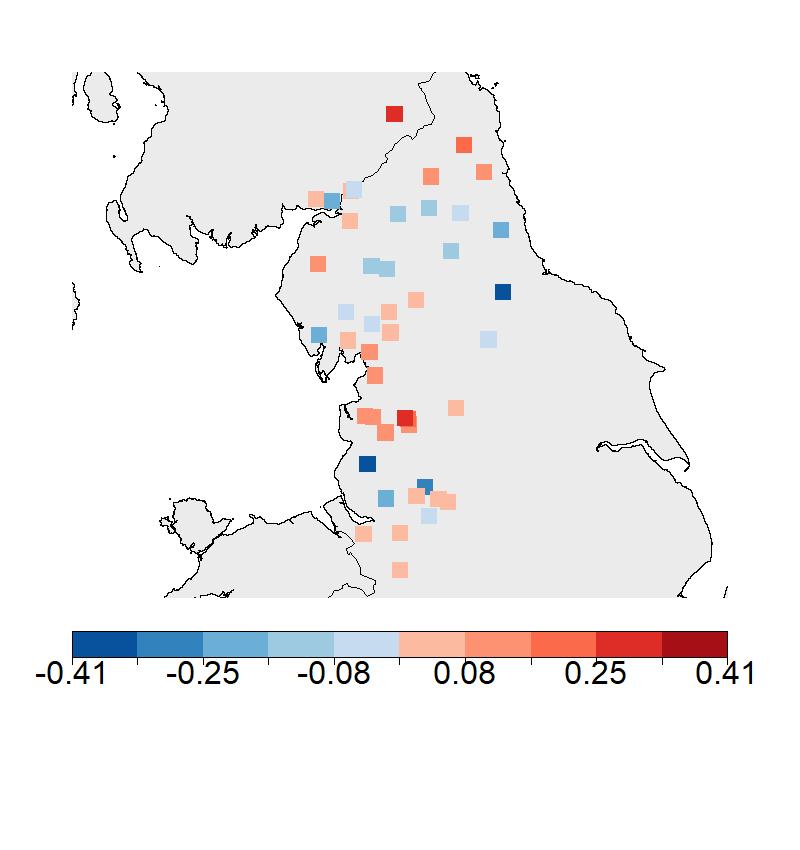}
\includegraphics[width=0.328\textwidth]{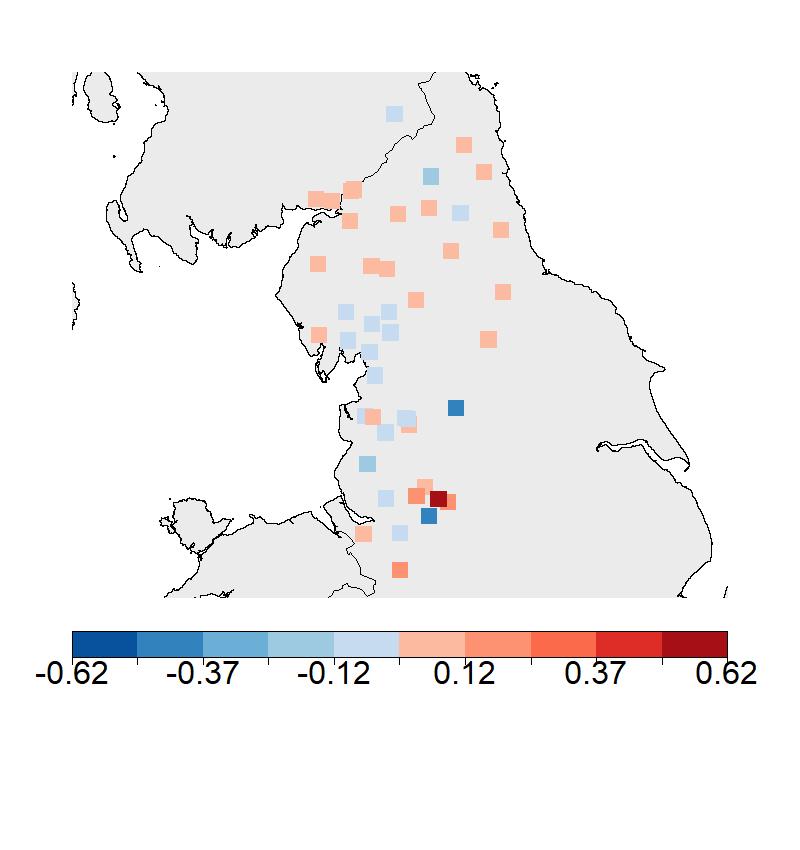}\vspace{-0.8cm}
\caption{Spatial illustration of the first twelve eigenvectors of the TPDM. The first row corresponds to the eigenvectors $\hat{\mathbf{U}}_{\cdot,1}$, $\hat{\mathbf{U}}_{\cdot,2}$ and $\hat{\mathbf{U}}_{\cdot,3}$ (left--right). Colour scales for eigenvectors are balanced such that red colours indicate positive values, while blue colours correspond to negative values; however, each plot has its own scale.}
\label{fig:U}
\end{figure}

We compute the eigendecomposition of $\hat{\Sigma}$ and obtain matrices $\hat{\mathbf{U}}$ and $\hat{\mathbf{D}}$, such that $\hat{\mathbf{U}}\hat{\mathbf{D}}\hat{\mathbf{U}}^{\mathrm{T}}=\hat{\Sigma}$. By investigating the spatial structure of the leading eigenvectors sequentially, we derive features of the extremal dependence across gauges. Figure~\ref{fig:U} shows the eigenvectors $\hat{\mathbf{U}}_{\cdot,1}, \ldots,\hat{\mathbf{U}}_{\cdot,12}$, plotted according to each gauge’s spatial location. The first eigenvector $\hat{\mathbf{U}}_{\cdot,1}$ has only positive values and accounts for the overall magnitude of extreme river flow events, with the highest values in the centre of the study area. The second eigenvector $\hat{\mathbf{U}}_{\cdot,2}$ shows a north-south divide, indicating that extreme river flow events tend to affect either the northern or the southern half of the study area more severely; this north-south divide corresponds with the topographical features described in Section~\ref{sec:Data}. Next, $\hat{\mathbf{U}}_{\cdot,3}$ shows a linear trend from the west to the east coast, with the exception of the most southern gauges, corresponding to the gauge's exposure to weather fronts from a westerly or south-westerly direction; the north-eastern gauges are protected by the Pennines, and the most southern gauges lie in the rain shadow of Snowdonia in northern Wales. The fourth eigenvector~$\hat{\mathbf{U}}_{\cdot,4}$ has a similar structure to $\hat{\mathbf{U}}_{\cdot,3}$, but it also splits the gauges close to the west coast based on their topology. Eigenvector $\hat{\mathbf{U}}_{\cdot,5}$ shows a general west-east trend, with the biggest differences being visible for the most southern gauges, and the spatial structure of $\hat{\mathbf{U}}_{\cdot,6}$ resembles that of Storm Desmond in Figure~\ref{fig:Locations} for the northern gauges. Some spatial structure is visible for $\hat{\mathbf{U}}_{\cdot,7}$, but it does not correspond to any known climatology. Finally, $\hat{\mathbf{U}}_{\cdot,8},\ldots,\hat{\mathbf{U}}_{\cdot,12}$ have no clear spatial structure and these eigenvectors may represent local variations between gauges. In summary, $\hat{\mathbf{U}}_{\cdot,1},\ldots,\hat{\mathbf{U}}_{\cdot,5}$, and potentially $\hat{\mathbf{U}}_{\cdot,6}$ and $\hat{\mathbf{U}}_{\cdot,7}$, appear to capture the large-scale spatial structure of extreme river flow events, while the remaining eigenvectors represent local dynamics.

\begin{figure}
\centering
\includegraphics[width=\textwidth]{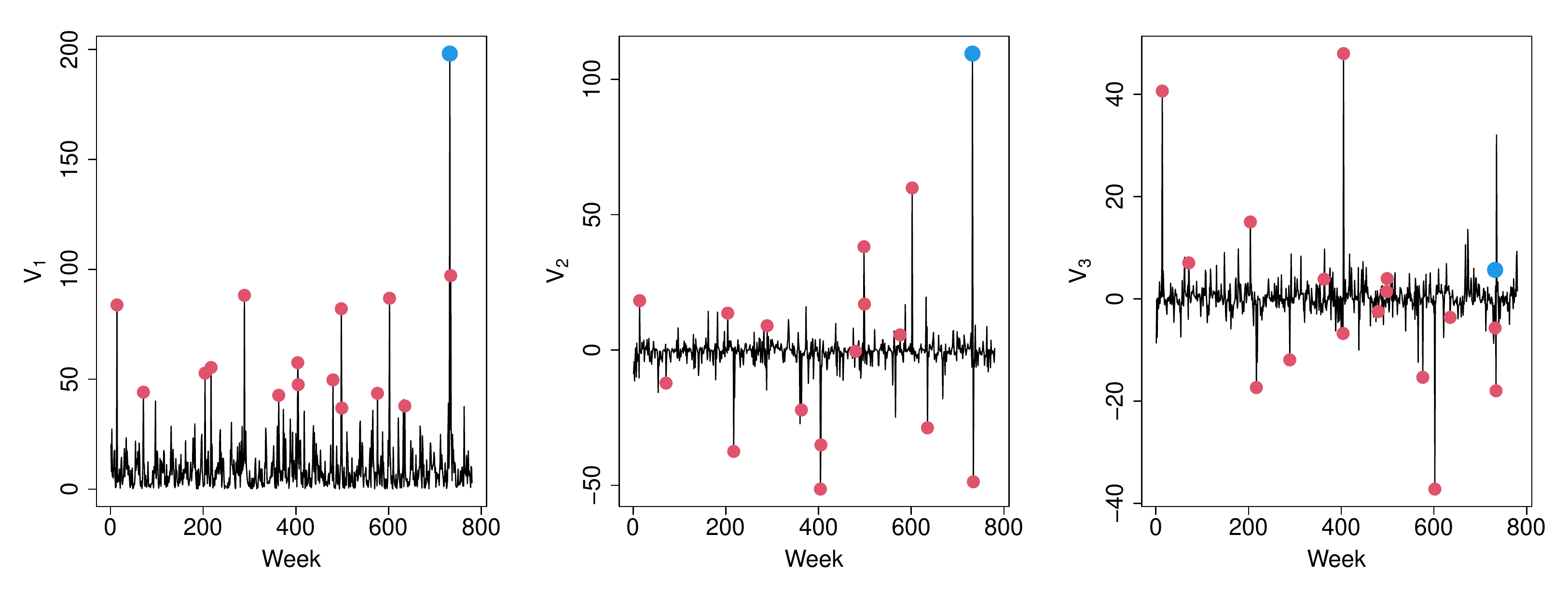}
\caption{Time series plots for the first three extremal principal components: $V_1$~(left), $V_2$~(middle) and $V_3$~(right). The red dots highlight the weeks for which $r_t$ exceeds the 98\% quantile of $\{r_t:t\in\mathcal{T}\}$. The blue dot highlights the extremal principal components related to Storm Desmond.}
\label{fig:TimeSeriesV}
\end{figure}

We conclude the analysis by studying the extremal principal components $\mathbf{v}_t = \hat{\mathbf{U}}^{\mathrm{T}} t^{-1}(\tilde{\mathbf{x}}_t)$ ($t\in\mathcal{T}^*$) defined in \eqref{eq:V}. The eigenvalues $\hat\lambda_1,\ldots,\hat\lambda_K$ relate to the scale of the extremal principal components, and Figure~\ref{fig:TPDM} right panel shows that they are quite small after the first six or seven components, with
$(\hat\lambda_1,\ldots,\hat\lambda_6) = (28.9, 4.7, 2.2, 1.7, 1.3, 1.2)$. Figure~\ref{fig:TimeSeriesV} shows time series plots for the extremal principal components associated to the first three eigenvectors. For Storm Desmond in 2015 (highlighted in blue), the first component indicates that this event caused extreme river flow across the region, the second extremal principal component contains the information that the most severe river flow was observed across the northern half of the study area, and the moderately positive value of the third extremal principal component shows that river flows were slightly more severe in the west than in the east (after accounting for the effects already described by the first two components). This agrees with the findings shown in Figure~\ref{fig:Locations} right panel. We find further agreements between high values for the first extremal principal components and recorded extreme river flow events. For instance, the highest value for the third extremal principal component relates to the highest observed levels for the southern gauges over the study period, which were caused by record levels of precipitation in the autumn of 2000.  

\section{Generating hazard event sets}
\label{sec:Generation}

\subsection{Introduction}

The generation of flood event sets requires accurate sampling from the upper tail of a $K$-dimensional random vector $\mathbf{X}$, with $K=45$ in Section~\ref{sec:Analysis}. Instead of $\mathbf{X}$, we consider sampling from the tail distribution of the regular-varying random vector $\tilde{\mathbf{X}}$ obtained by the marginal transformation~\eqref{eq:Transform}. One could generate values for $\tilde{\mathbf{X}}$ based on a fitted tail dependence model, such as \citet{Tawn1988} or \cite{Carvalho2014}. However, existing multivariate extreme value models are ill-equipped to handle the dimension of the considered river flow application, due to the complexity of the angular measure $H_X(\cdot)$ and the low number of observed extreme events.

Our proposed generative framework for hazard event sets is based on sampling values for the extremal principal components $\mathbf{V}$. By applying the inverse of the transformations \eqref{eq:V} and \eqref{eq:Transform} to the generated values of $\mathbf{V}$, we obtain samples for the random vector~$\mathbf{X}$ of interest. Since the extremes of~$\tilde{\mathbf{X}}$~and~$\mathbf{V}$ are linked, an accurate model for the tail of $\mathbf{V}$ will capture the extremal dependence of the components of $\tilde{\mathbf{X}}$. As $\mathbf{V}$ is regularly-varying with index~$\alpha=2$ (Section~\ref{sec:Cooley2019}) and $\mathbb{P}\left(||\mathbf{V}||_2 \leq r \right) = \exp\left[-(r/K)^{-2}\right]$~($r>0$),we only need to estimate $H_V(\cdot)$ which describes the behavior of
$\mathbf{W} =  \left.\frac{\mathbf{V}}{||\mathbf{V}||_2}\,\right|\,||\mathbf{V}||_2 > r_V $
as $r_V\to\infty$. We assume the limit holds above a sufficiently high~$r_V$, which we set to the empirical 94\% quantile of $||\mathbf{V}||_2$, and we denote $\{\mathbf{w}_i\in\mathbb{S}^{K-1}\,:\, i=1,\ldots,n\}$ as the observations for $\mathbf{W}$.

\begin{figure}
\centering
\includegraphics[width=0.3\textwidth]{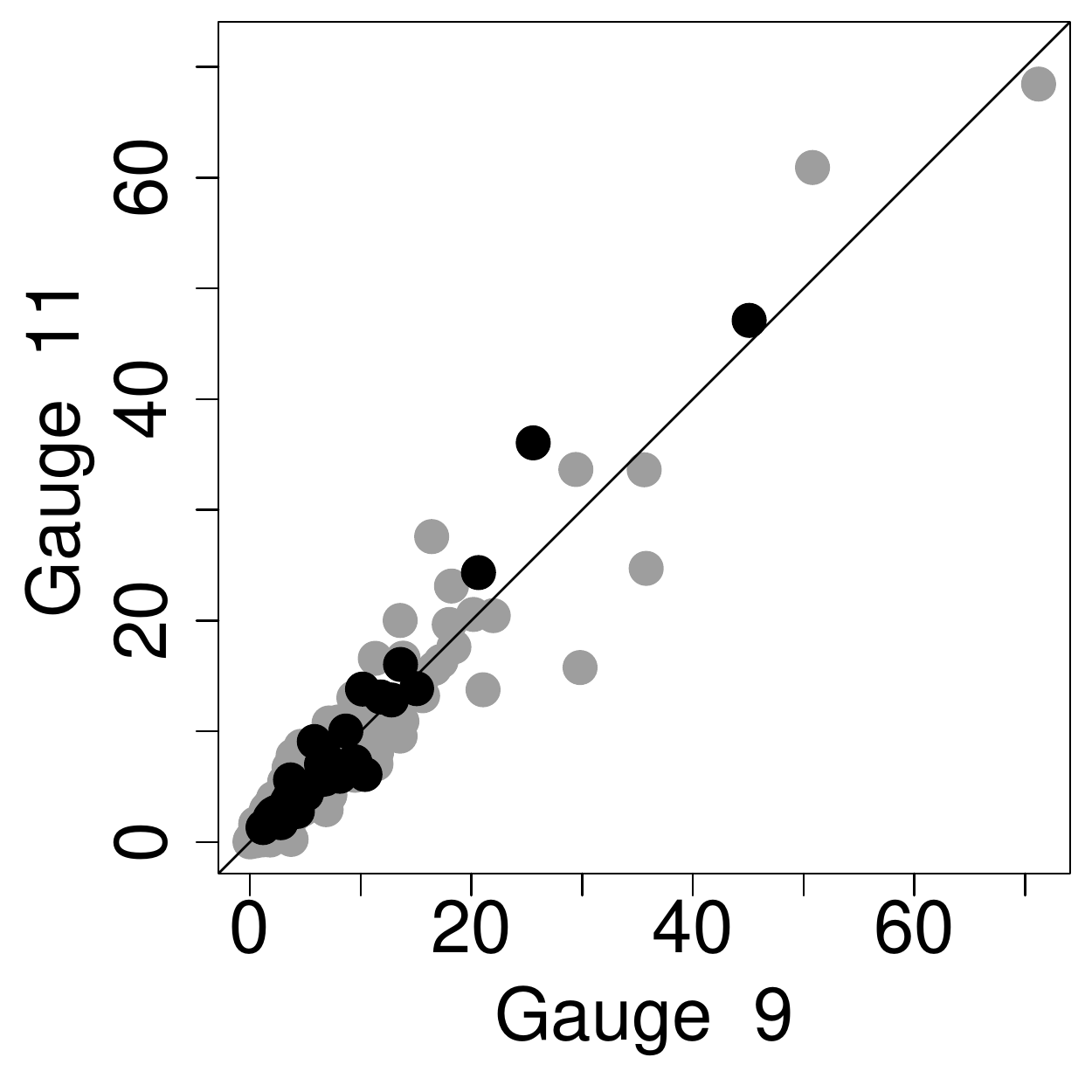}\hspace{0.4cm}
\includegraphics[width=0.3\textwidth]{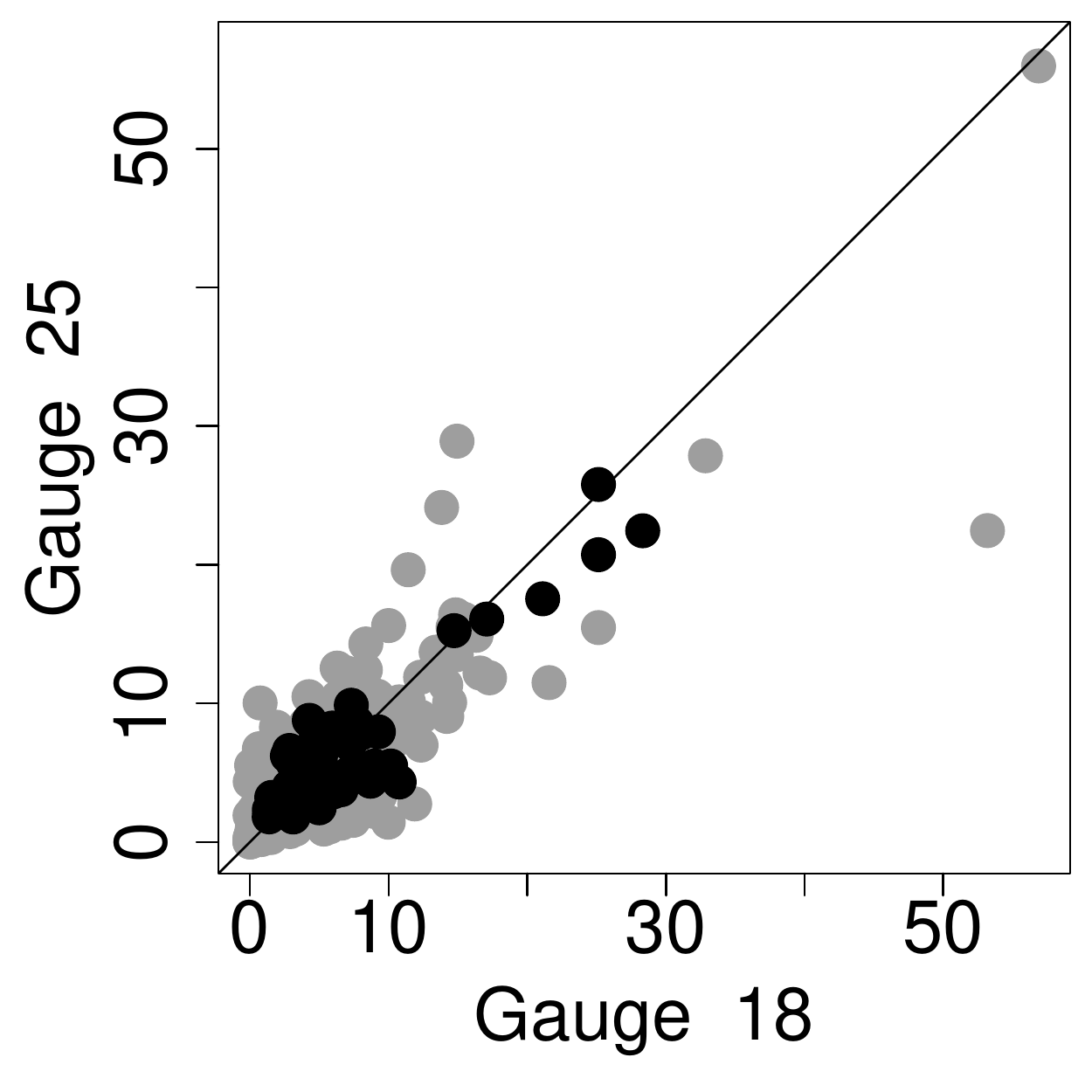}\hspace{0.4cm}
\includegraphics[width=0.3\textwidth]{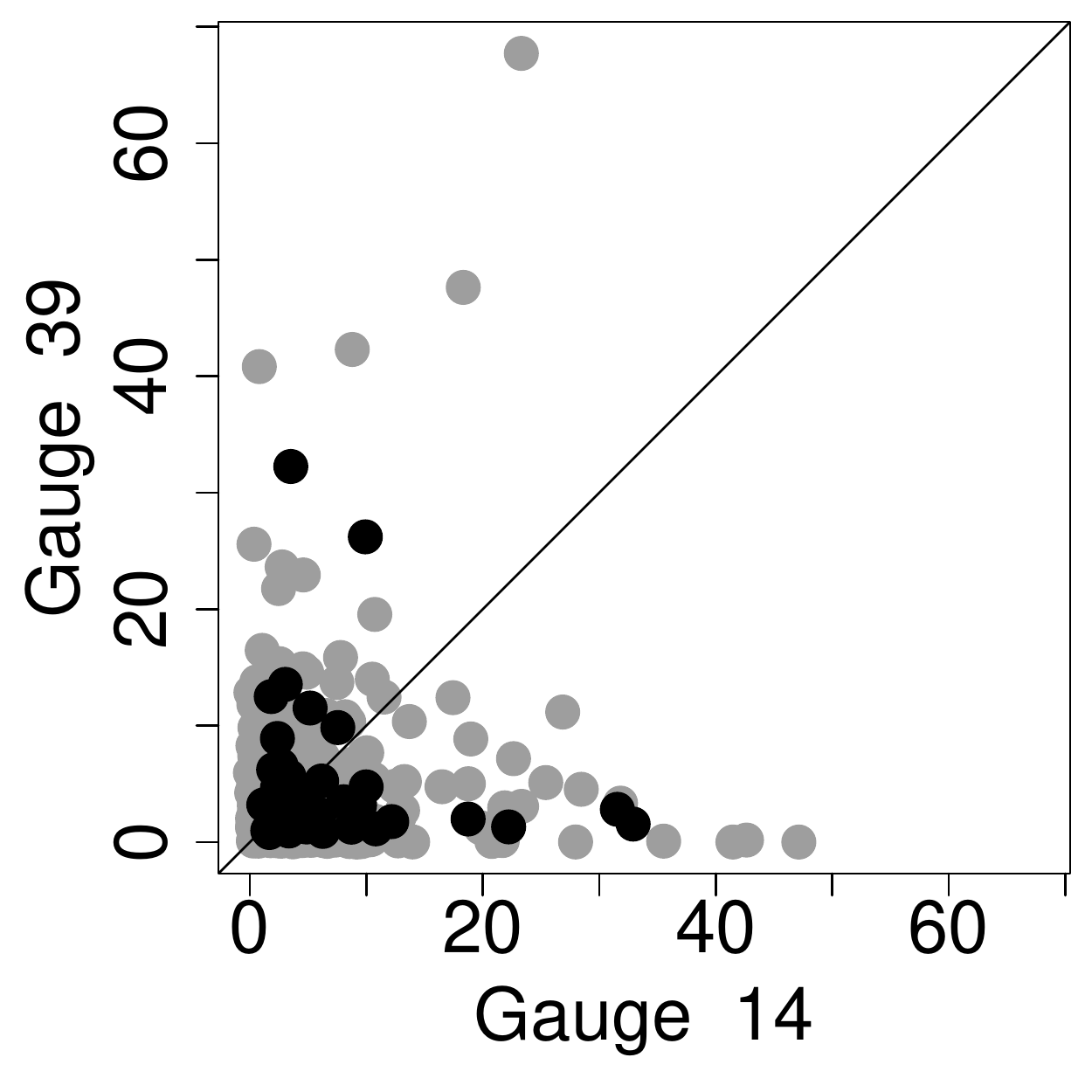}
\caption{Simulated (grey) and observed (black) extreme events of $\tilde{\mathbf{X}}$ for three pairs of gauges. The generated extreme events are based on a sample of size 4,400, corresponding to a 200-year hazard event set, while the observed extremes are for the period 1980-2018. Samples are generated using the framework described in Sections~\ref{sec:W} and \ref{sec:Selectingm}. The two gauges in the left panel are flow-connected and spatially close, the gauges in the middle panel are spatially close but not flow-connected, and the gauges in the right panel are neither spatially close nor flow-connected.}
\label{fig:ResultsPairs}
\end{figure}

At first glance, the estimation of $H_V(\cdot)$ is as difficult as that of $H_X(\cdot)$. The key difference is that the components of $\mathbf{V}$ have diminishing contributions to the extremes of $\tilde{\mathbf{X}}$. For the analysis in Section~\ref{sec:PCAResults}, we find $\lambda_1+\cdots+\lambda_6=39.75$, compared to $\lambda_{7}+\cdots+\lambda_{K}=5.25$, indicating that the first six eigenvalue / eigenvector pairs explain about 88\% of the scale in the extremes. This observation motivates our proposal to estimate the distribution of $\mathbf{W}$ as a combination of (i)~a flexible model capturing the full dependence structure in the first $m$ components and (ii) a simplified framework for the remaining components. In Section~\ref{sec:W}, we describe our models for parts (i) and (ii) for fixed~$m$. With our models in place, we are better able to explore the choice of $m$ in Section~\ref{sec:Selectingm}. Figure~\ref{fig:ResultsPairs} previews that samples generated via our framework exhibit a good agreement in the extremal dependence structure of sampled and observed values of $\tilde{\mathbf{X}}$ for three pairs of gauges (from left to right: strongly dependent, moderately dependent and weakly dependent). Model uncertainty is considered in Section~\ref{sec:Uncertainty} and an analysis of the generated hazard event sets is performed in Section~\ref{sec:Results}. 
 
\subsection{Generative framework and sampling algorithm}
\label{sec:W}

We first describe separate potential models for the random vectors $\mathbf{W}_{1:m}=(W_1,\ldots,W_m)$ and $\mathbf{W}_{(m+1):K}=(W_{m+1},\ldots,W_K)$, but then transition to a joint modelling framework which achieves superior results.

\paragraph{Modelling the first $m$ components}

We require a sufficiently flexible model to capture the potentially complex dependence structure of the components in~$\mathbf{W}_{1:m}$. Consider the random vector $\mathbf{W}_{1:m}\,/\,||\mathbf{W}_{1:m}||_2$. Given the observations $\left\{ \mathbf{w}_{i,1:m}\,:\,i=1,\ldots,n\right\}$ for $\mathbf{W}_{1,m}$, we model the distribution of $ \mathbf{W}_{1:m}\,/\,||\mathbf{W}_{1:m}||_2$ using a kernel density estimate appropriate for spherical data. As suggested by \cite{Hall1987}, our kernel of choice is the von Mises-Fisher distribution with density 
$
h(\mathbf{z}; \bm\mu, \kappa) = c_0(\kappa) \exp\left( \kappa \mathbf{z}^{\mathrm{T}}\bm\mu\right) 
$
for $\mathbf{z}\in\mathbb{S}^{m-1}$, 
where $\bm\mu\in\mathbb{S}^{m-1}$ and $\kappa>0$ are termed the mean direction and concentration parameter, respectively, and $c_0(\kappa)$ is a normalizing constant. The kernel density estimate for $ \mathbf{W}_{1:m}\,/\,||\mathbf{W}_{1:m}||_2$ is then given by
\begin{equation}
\hat{h}_V(\mathbf{z};\kappa)~=~\frac{1}{n}\, \, \sum_{i=1}^{n} h\left(\mathbf{z}; \frac{\mathbf{w}_{i,1:m}}{||\mathbf{w}_{i,1:m}||_2}\,,\, \kappa\right) \qquad (\mathbf{z}\in\mathbb{S}^{m-1}).
\label{eq:MisesFisher}
\end{equation}
We estimate $\kappa$ using the \texttt{vmf.kde.tune()} function in the \texttt{R} package \texttt{Directional}; the function obtains an estimate using maximum likelihood inference. The robustness of this approach for our analysis was verified using bootstrapping, and only small variations in the estimated $\kappa$ were found.

\paragraph{Modelling the remaining components} An overly simple approach sets the components $W_{m+1},\ldots,W_K$ to their empirical average, and only samples values for~$W_1,\ldots,W_m$ using the kernel density estimate \eqref{eq:MisesFisher}. However, generated flood event sets exhibited a higher degree of extremal dependence than we observed in the UK river flow data. An alternative approach samples values for $W_{m+1},\ldots,W_K$ from their joint empirical distribution function. For the UK river flow, this sampling strategy improved upon the first one, but the generated extremes for spatially distant gauges appeared more dependent than in reality. The limitation of both approaches is that they assume $\mathbf{W}_{1:m}$ and $\mathbf{W}_{{m+1}:K}$ to be independent, ignoring the fact that $V_i$ and $V_j$~($i,j=1,\ldots,K$) are asymptotically dependent (Section~\ref{sec:Cooley2019}). Since separate modelling of $\mathbf{W}_{1:m}$ and $\mathbf{W}_{(m+1):K}$ is unable to capture all properties of the underlying process, we instead propose a joint modelling framework.

\paragraph{Joint modelling framework} We start by transforming $\mathbf{W}$ into a random variable $\mathbf{Z}$ on the ($m+1$)-dimensional unit sphere $\mathbb{S}^m$. The idea is that the first $m$ components of $\mathbf{Z}$ represent the information contained in $\mathbf{W}_{1:m}$, while the final component $Z_{m+1}$ summarizes some aspects of the random vector $\mathbf{W}_{(m+1):K}$. Let $\mathbf{Z}=(Z_1,\ldots,Z_{m+1})$ with $Z_j = W_j$~($j=1,\ldots,m$) and
\begin{equation}
Z_{m+1} =
\begin{cases}
~~\sqrt{1-\sum_{j=1}^m W_j^2} & \quad\mbox{if}~W_{m+1}\geq 0,\\
-\sqrt{1-\sum_{j=1}^m W_j^2} & \quad\mbox{if}~W_{m+1}<0.
\end{cases}
\label{eq:Z}
\end{equation}
Given observations $\{\mathbf{z}_i: i=1,\ldots,n\}$, we obtain the kernel density estimate
\begin{equation}
\hat{h}_Z(\mathbf{z},\kappa_Z) =
\frac{1}{n}\sum_{i=1}^{n} h\left(\mathbf{z}; \mathbf{z}_{i}, \kappa_Z\right) \qquad (\mathbf{z}\in\mathbb{S}^m).
\label{eq:ZKDE}
\end{equation}
In addition to the ability to fully capture the dependence in $W_1,\ldots,W_m$, this model also accounts for some of the dependence between $\mathbf{W}_{1:m}$ and $\mathbf{W}_{(m+1):K}$ at the cost of one additional dimension.

The next step is to define a mapping which converts a sample $\mathbf{z}^*$ for $\mathbf{Z}$, obtained by drawing from the kernel density estimate in \eqref{eq:ZKDE}, into a sample $\mathbf{w}^*$ for $\mathbf{W}$. Based on the defined connection between $\mathbf{W}$ and $\mathbf{Z}$, the first $m$ components of $\mathbf{w}^*$ are set to those of $\mathbf{z}^*$, i.e., $w_j^*=z_j^*~(j=1,\ldots,m)$. Consequently, we are left with transforming the sampled value $z^*_{m+1}$ into a sample $\mathbf{w}^*_{(m+1):K}$ for $\mathbf{W}_{(m+1):K}$, while accounting for the dependence amongst $W_1,\ldots,W_K$.

We propose to use the observations of $\mathbf{Z}$ most similar to the sampled vector $\mathbf{z}^*$ in order to identify possible patterns for $w^*_1,\ldots,w^*_K$, conditional on $w_j^*=z_j^*~(j=1,\ldots,m)$. We propose a nearest-neighbour approach, with the index $q$ of the observation closest to the sample given by
\begin{equation}
q ~=~ \argmax_{i=1,\ldots,n}~ \mathbf{z}_i ^{\mathrm{T}} \mathbf{z}^*.
\label{eq:NN}
\end{equation}
The sample $\mathbf{w}^*_{(m+1):K}$ is then set to a scaled version of $\mathbf{w}_{q,(m+1):K}$, yielding
\begin{equation}
\mathbf{w}^* = \left(z_1^*,\ldots,z_m^*, \left|\frac{z_{m+1}^*}{z_{q,m+1}}\right| w_{q,m+1}, \ldots,\left|\frac{z_{m+1}^*}{z_{q,m+1}}\right| w_{q,K}\right).
\label{eq:JM}
\end{equation}
Note, the described mapping yields $\mathbf{w}^* = \mathbf{w}_i$ for $\mathbf{z}^*=\mathbf{z}_i$~($i=1,\ldots,n$).

\begin{figure}
\centering
\includegraphics[width=0.325\textwidth]{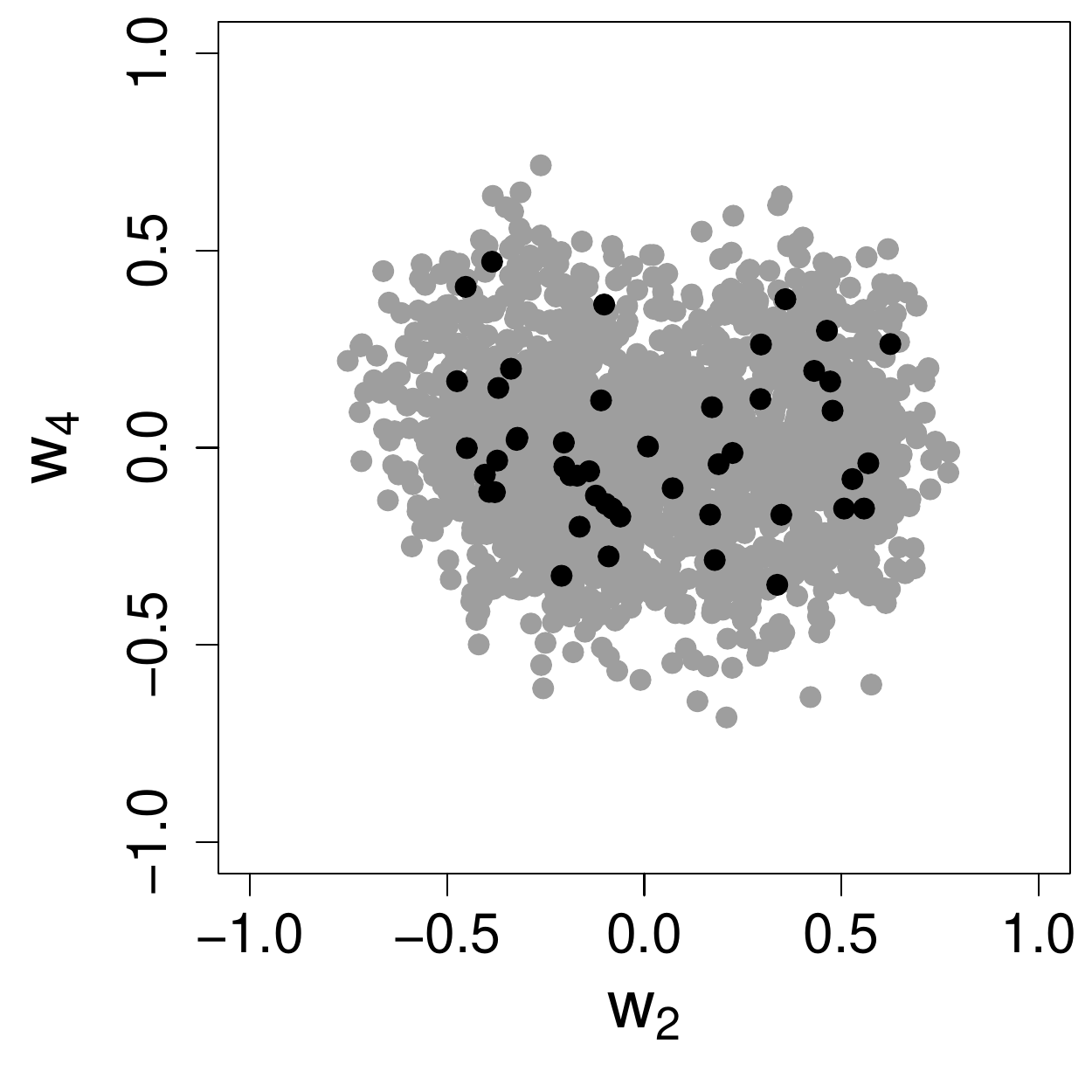}
\includegraphics[width=0.325\textwidth]{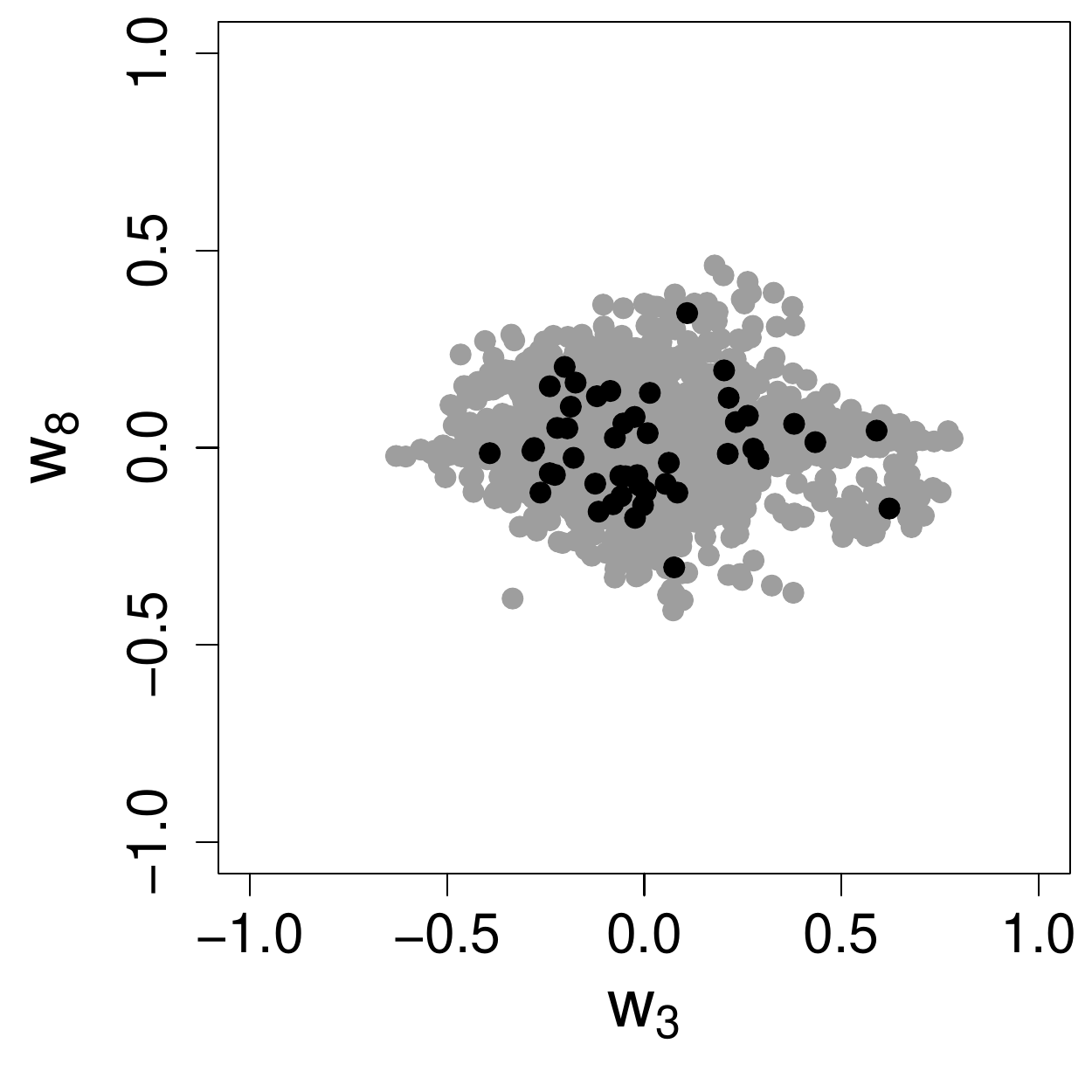}
\includegraphics[width=0.325\textwidth]{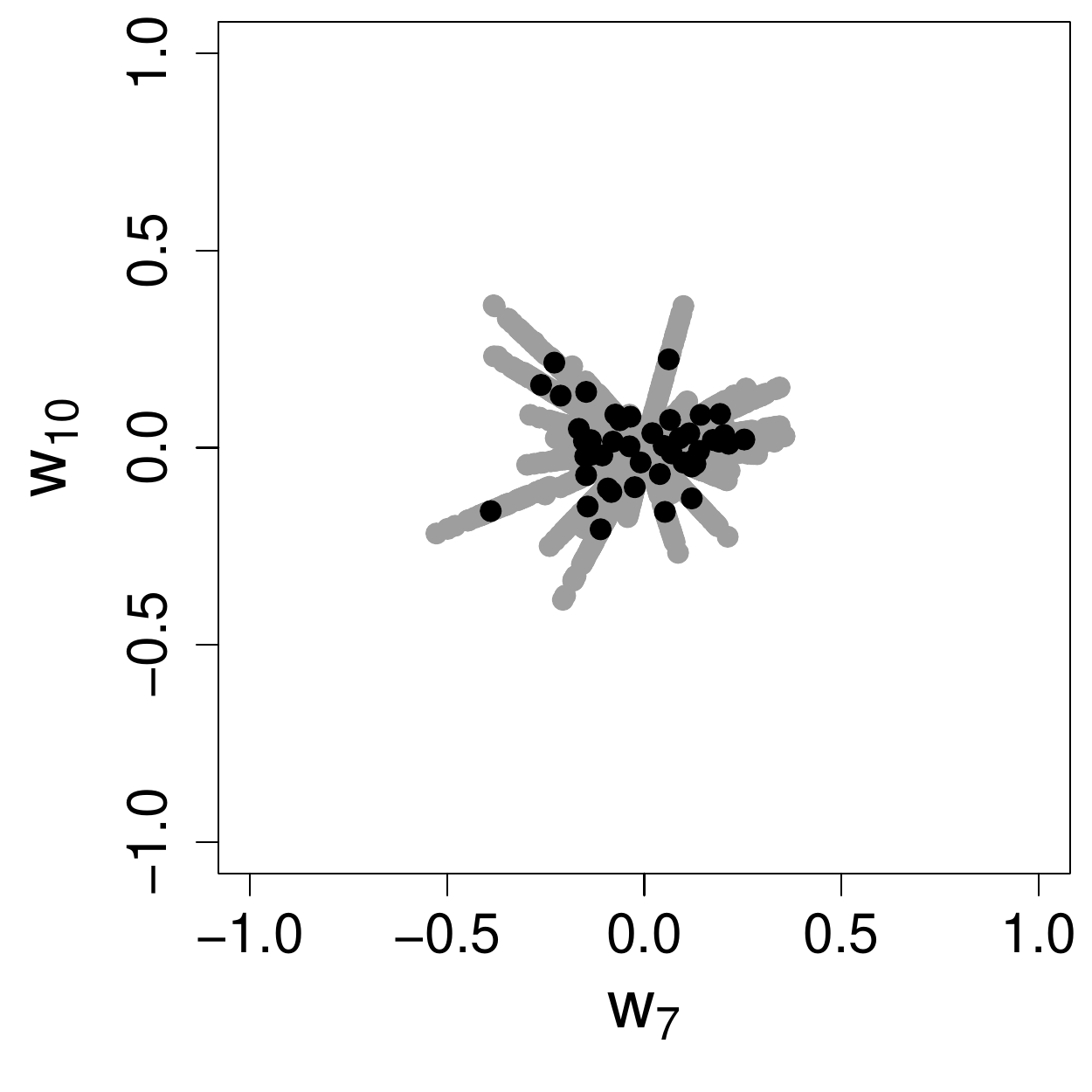}
\caption{Pairwise plots of generated (grey) and observed (black) values for $(W_2,W_4)$ (left), $(W_3,W_8)$ (middle) and $(W_7,W_{10})$ (right) for the UK river flow data. The number of observed values is $n=47$, and 2,000 samples were generated using the procedure outlined in expressions~\eqref{eq:ZKDE}, \eqref{eq:NN} and \eqref{eq:JM}. The parameter $m$ was set to $m=6$. As such, the number of components directly modelled via the density in \eqref{eq:ZKDE} is two (left), one (middle) and zero~(right). The plots also illustrate that the variance of $W_j$ tends to decrease with increasing index $j$~($j=1,\ldots,K$), which reflects the difference in scale of the extremal principal components~$V_1,\ldots,V_K$.}
\label{fig:SamplesW}
\end{figure}

Figure~\ref{fig:SamplesW} shows that the pairwise dependence of sampled and observed values for $\mathbf{W}$ generally agree. Here, $m$ was set to the preliminary value of $m=6$. The discrete nature of the right panel in Figure~\ref{fig:SamplesW} is due to there only being $n$ possible pairwise patterns for the sampled values $\mathbf{w}^*_{(m+1):K}$. Since the components of $\mathbf{W}_{1:m}$ contain most of the information on the tail distribution of $\tilde{\mathbf{X}}$, this restrictive sampling of $\mathbf{w}^*_{(m+1):K}$ will only have a small effect on the samples for $\tilde{\mathbf{X}}$. Furthermore, the discrete nature of $\mathbf{w}^*_{(m+1):K}$ will be masked after accounting for uncertainty in the estimates of the marginal distributions and the TPDM in Section~\ref{sec:Uncertainty}.

\paragraph{Simulating values for $\tilde{\mathbf{X}}$} Using the framework described above, samples for $\tilde{\mathbf{X}}$ are generated as follows: 
\begin{enumerate}
\item Obtain a sample $\mathbf{z}^*\in\mathbb{S}^m$ from the distribution with estimated density \eqref{eq:ZKDE}.
\item Extract the index $q$ in \eqref{eq:NN} and derive the angular component $\mathbf{w}^*\in\mathbb{S}^{K-1}$ using \eqref{eq:JM}.
\item Sample $r^*$ from a Fr{\'e}chet distribution with $\mathbb{P}\left(||\mathbf{V}||_2 \leq r \right) = \exp\left[-(r/K)^{-2}\right]$.
\item Calculate the generated sample $\mathbf{v}^*$ for $\mathbf{V}$, $\mathbf{v}^* = (v_1^*,\ldots,v_K^*) = (r^*w^*_1,\ldots,r^*w^*_K)\in\mathbb{R}^K$.
\item Apply the inverse of \eqref{eq:V} to obtain a sample $\tilde{\mathbf{x}}^*\in\mathbb{R}^{K}_+$ for $\tilde{\mathbf{X}}$,
\[
\tilde{\mathbf{x}}^* = \tau \left\{ \sum_{j=1}^K v_j^* \,\mathbf{U}_{ \cdot,j}\right\},
\]
where $\tau(\cdot)=\log\left[\exp(\cdot) +1 \right]$ is applied component-wise, and $\mathbf{U}_{\cdot,j}$ is the $j$-th eigenvector of the estimated TPDM. Note, $\tilde{\mathbf{x}}^*\in\mathbb{R}^{K}_+$, while, in steps 1,2, and 4, the sampled values can be positive or negative.
\end{enumerate}

The nearest-neighbour approach in \eqref{eq:NN} and \eqref{eq:JM} is dimension-reducing: the sample $\mathbf{w}^*$ for~$\mathbf{W}$ lies within a $m$-dimensional non-connected manifold on $\mathbb{S}^{K-1}$. Further, as $n\to\infty$, the sampling space of $\mathbf{w}^*$ is dense in the support of $H_V(\cdot)$, i.e., there exists a point in the sampling space within any neighbourhood of a possible outcome of~$\mathbf{W}$.

We can also consider alternatives to the nearest-neighbour approach. One possible approach is to sample the index $q$ amongst the observations of $\mathbf{Z}$ reasonably close to $\mathbf{z}^*$. Another alternative is to sample $\mathbf{w}^*_{(m+1):K}$ from a von Mises-Fisher distribution with mean direction $\mathbf{w}_{q,(m+1):K}$. Both of these approaches achieve convergence of the sampling distribution to $H_V(\cdot)$ as $n\to\infty$, and perform similar to the nearest-neighbour approach. However, they require more fine-tuning in practice: selection of the sampling distribution for $q$, or a value for the concentration parameter in the von Mises-Fisher distribution.

\subsection{Selection of \texorpdfstring{$m$}{m}}
\label{sec:Selectingm}

The generative framework in Section~\ref{sec:W} requires the selection of $m$, which balances two competing aspects. For a small value of $m$, the distribution of the random vector $\mathbf{Z}$ in \eqref{eq:Z} can be estimated to a reasonable degree of accuracy, but we may underestimate the variance of $\mathbf{W}$ due to the restrictive sampling approach for $\mathbf{w}^*_{(m+1):k}$. A large $m$ provides greater model flexibility, but the approach suffers from the curse of dimensionality: if $m$ increases, the relatively few extreme observations of $\mathbf{Z}$ become more isolated in $\mathbb{S}^m$.

In classical principal component analysis, multiple rules-of-thumb to select the number of principal components exist. One such rule is to look for an 'elbow' in the scree plot; the scree plot in Figure~\ref{fig:TPDM} indicates that the elbow lies between $m=5$ and $m=8$ for the UK river flow data. Another rule-of-thumb for PCA on correlation matrices is to keep all components with eigenvalue greater than 1. Due to data preprocessing, our TPDM has diagonal elements of 1, and applying this rule yields $m=6$. While these heuristics motivate a moderate value for $m$, we can also check whether certain types of extreme events are poorly explained by such a choice. We studied the extreme events which are poorly captured by the first $m=5,\ldots,8$ eigenvectors and we found no common spatial pattern of these events; our analysis is provided in the supplementary material.

\citet{Drees2021} propose a graphical diagnostic, tailored towards the analysis of extreme values, for the selection of $m$. The diagnostic in \citet{Drees2021} arises from the assumption that the angular measure is concentrated on a lower-dimension linear subspace of $\mathbb{S}^{K-1}$, and the diagnostic measures the risk distance from the lower-dimensional projection. Their diagnostic performs well in simulations where this assumption is true, but the performance is less clear when this only holds approximately. Since the first twelve eigenvalues in Figure~\ref{fig:TPDM} right panel are not zero, the diagnostic by \citet{Drees2021} would require the user to select $m$ when the assessed risk is close enough to zero; a choice similar to using the aforementioned scree plot or the reconstruction error analysed in the supplementary material.

\begin{figure}
\centering
\includegraphics[width=0.5\textwidth]{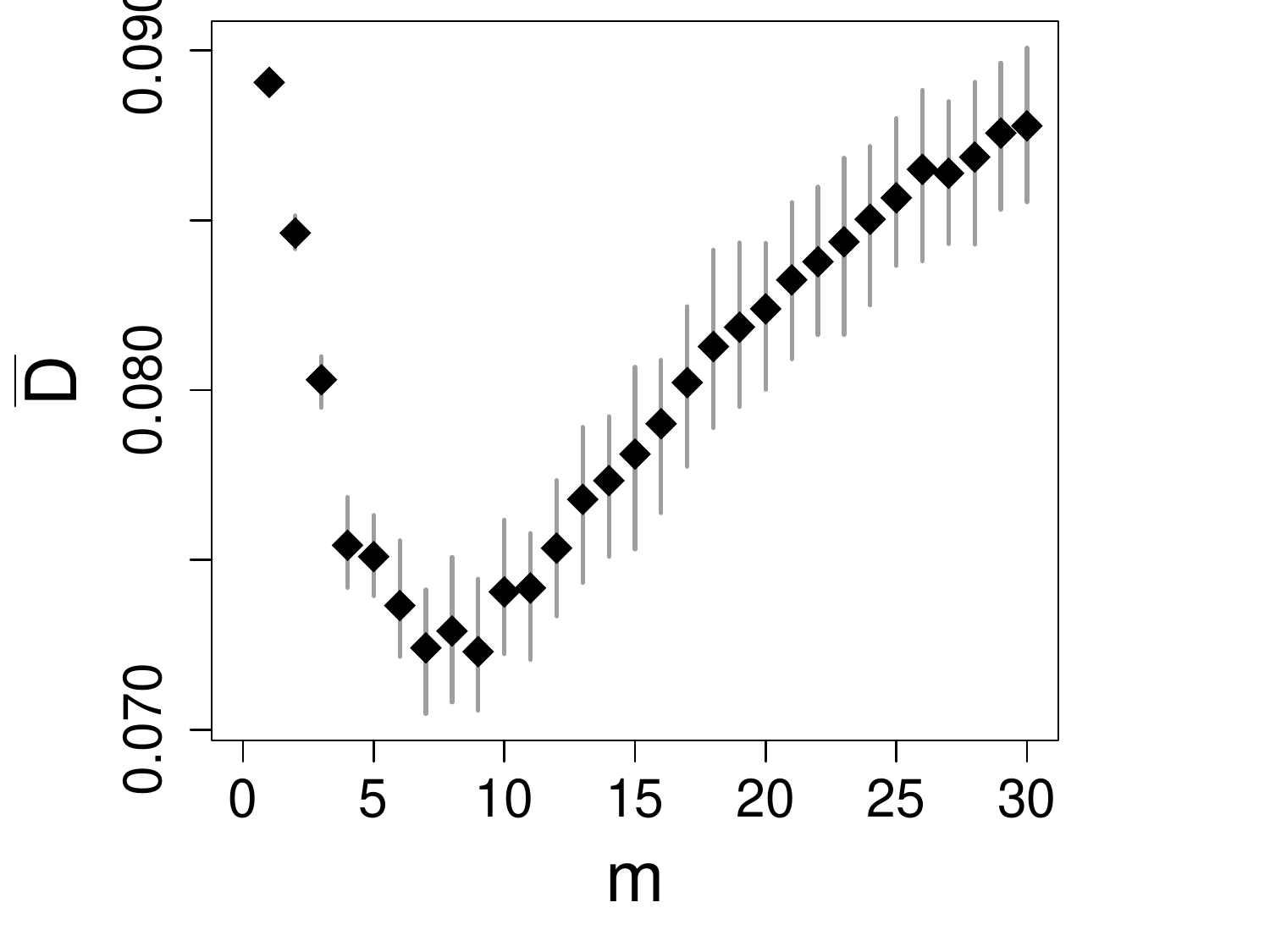}
\caption{Values of $\bar{D}$ for $m=1,\ldots,30$ estimated based on 100 leave-one-out cross validations for the UK river flow data. The grey lines correspond to the central 90\% confidence interval of $\bar{D}$ of the different values of $m$.}
\label{fig:Selectionofm}
\end{figure}

We propose selecting $m$ by leave-one-out cross validation. Unlike the scree plot, this not only accounts for how well modelling the leading $m$ eigenvalues captures dependence, but also accounts for how well the nearest neighbour approach accounts for any residual behaviour. For sample $i$, we remove the $i$-th extreme event $\tilde{\mathbf{x}}^{(i)}$~($i=1,\ldots,n$) from the data for $\tilde{\mathbf{X}}$ and obtain an estimate $\hat\Sigma^{(-i)}$ as in \eqref{eq:Sigmahat} using the remaining data, with the value $r_V$ being fixed to the 94\% empirical quantile across all samples. As such, each estimated TPDM $\hat\Sigma^{(-i)}$ is derived from $n-1=46$ extreme events in our river flow application. We then generate 2,000 samples $\tilde{\mathbf{x}}^{(-i)}_1, \ldots,\tilde{\mathbf{x}}^{(-i)}_{2000}$ for $\hat\Sigma^{(-i)}$ using the generative framework in Section~\ref{sec:W} for the different possible values of $m$. Performance is assessed based on the proximity of $\tilde{\mathbf{x}}^{(-i)}_1, \ldots,\tilde{\mathbf{x}}^{(-i)}_{2000}$ and the $i$-th removed extreme event $\tilde{\mathbf{x}}^{(i)}$ using 
\[
D_i = 1 - \max_{j=1,\ldots,2000} \left(\frac{\tilde{\mathbf{x}}^{(i)}}{\left|\left|\tilde{\mathbf{x}}^{(i)}\right|\right|} \,\bullet\, 
\frac{\tilde{\mathbf{x}}_j^{(-i)}}{\left|\left|\tilde{\mathbf{x}}_j^{(-i)}\right|\right|}\right),
\]
where $\bullet$ is the dot product and $D_i$ close to $0$ corresponds to close proximity of the samples and the removed extreme event. The optimal $m$ then minimizes the average error $\bar{D} = n^{-1}\left(D_1+\cdots+D_n\right)$. Figure~\ref{fig:Selectionofm} shows that $\bar{D}$ takes is minimum between $m=7$ and $m=9$, and we use $m=7$ in the remainder.

\subsection{Handling uncertainty}
\label{sec:Uncertainty}

So far, we have ignored any uncertainty in the marginal model estimates in Section~\ref{sec:Marginals} and the estimated TPDM $\hat{\Sigma}$ in \eqref{eq:Sigmahat}. Herein, we use nonparametric bootstrapping to account for both these sources of uncertainty in our generative framework in Section~\ref{sec:W}. We start by resampling the data for the random vector $\mathbf{X}$ with replacement, and by drawing values for the shape parameters $\xi_1,\ldots,\xi_K$ in \eqref{eq:GPD} from their joint posterior distribution obtained in Section~\ref{sec:Marginals}; one may also estimate the shape parameters for each new sample, but this approach was not pursued due to computational cost of the approach by \citet{RohrbeckTawn2020}. The GPD scale parameters $\sigma_1,\ldots,\sigma_K$ are then estimated as described in Section~\ref{sec:Marginals} using maximum likelihood estimation. By applying the marginal transformation \eqref{eq:Transform} and deriving the empirical estimate $\hat{\Sigma}$ \eqref{eq:Sigmahat}, we obtain a different TPDM for each sample.

Let $\hat\Sigma_i$ denote the estimated TPDM for the $i$-th resampled data set of $\mathbf{X}$. A sample $\mathbf{x}^*_i$ is then generated as described in Section~\ref{sec:W}, with a common value for $m$; we select $m=7$ for the UK river flow data. This approach also leads to greater variety of patterns of $\mathbf{w}_{(m+1):K}$ and, thus, reduces the possible limitations found in Figure~\ref{fig:SamplesW} right panel. However, this effect cannot be easily illustrated since the interpretation of the $j$-th eigenvectors~($j=1,\ldots,K$) may vary across the resampled data sets.

\section{Analysis of generated flood event sets}
\label{sec:Results}

\subsection{Validation and analysis}
\label{sec:Validation}

To validate our approach described in Section~\ref{sec:Generation}, we simulate 100 event sets of the same length as the original data. As such, each set comprises $T=848$ samples (Section~\ref{sec:Data}). Herein, we compare our generated values to the original river flow measurements, i.e., we analyse the performance of the samples for the random vector $\mathbf{X}$. We first assess the samples at a gauge level by comparing the fifty largest order statistics of the observed and simulated event sets. Our results show that the observed order statistics generally lie within their corresponding central 90\%  sampling intervals; quantile-quantile plots for three gauges are provided in the supplementary material. The marginal distributions of the sampled and observed values of $\mathbf{X}$ show agreement.

\begin{figure}[t]
\centering
\includegraphics[width=0.32\textwidth]{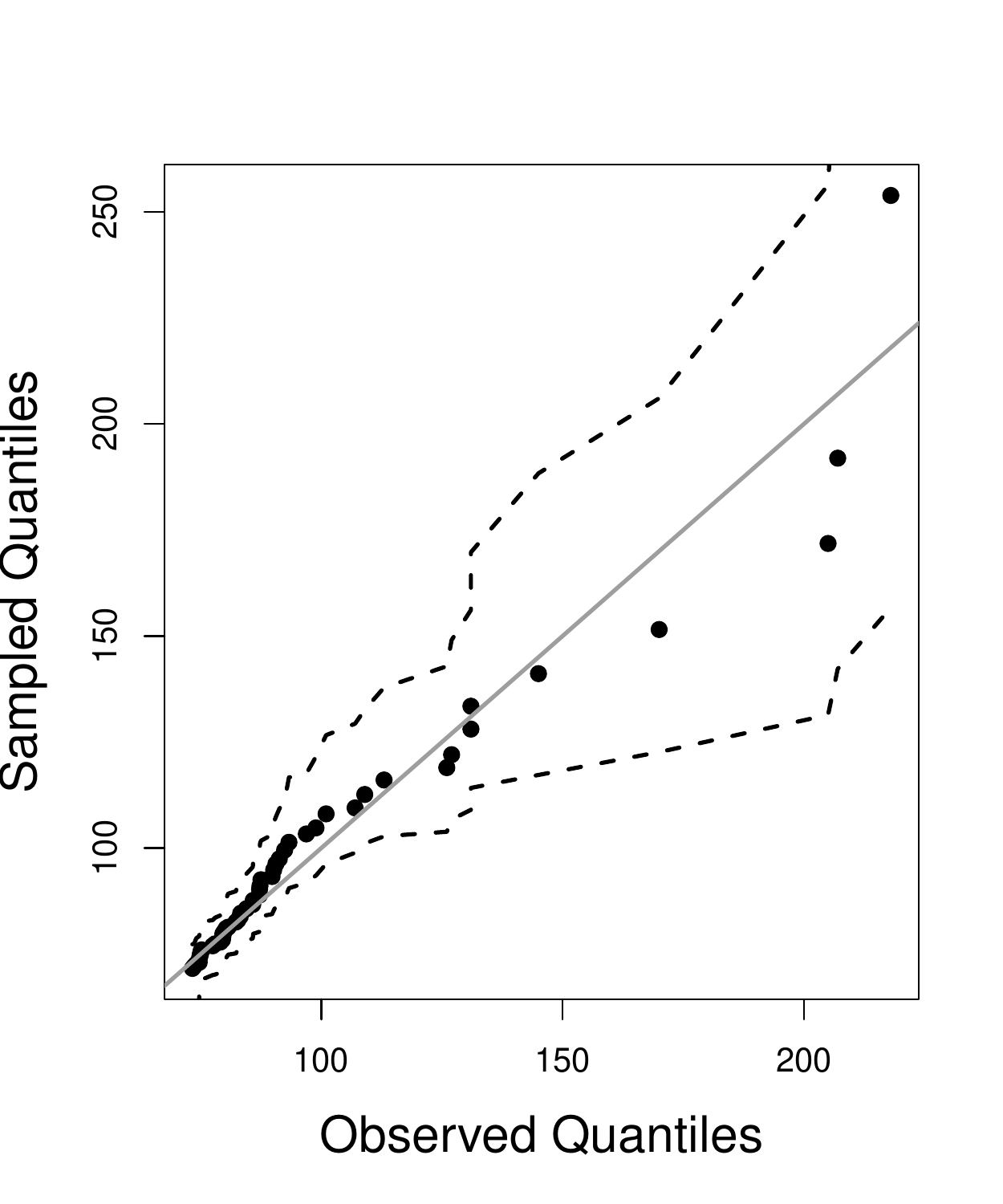}
\includegraphics[width=0.32\textwidth]{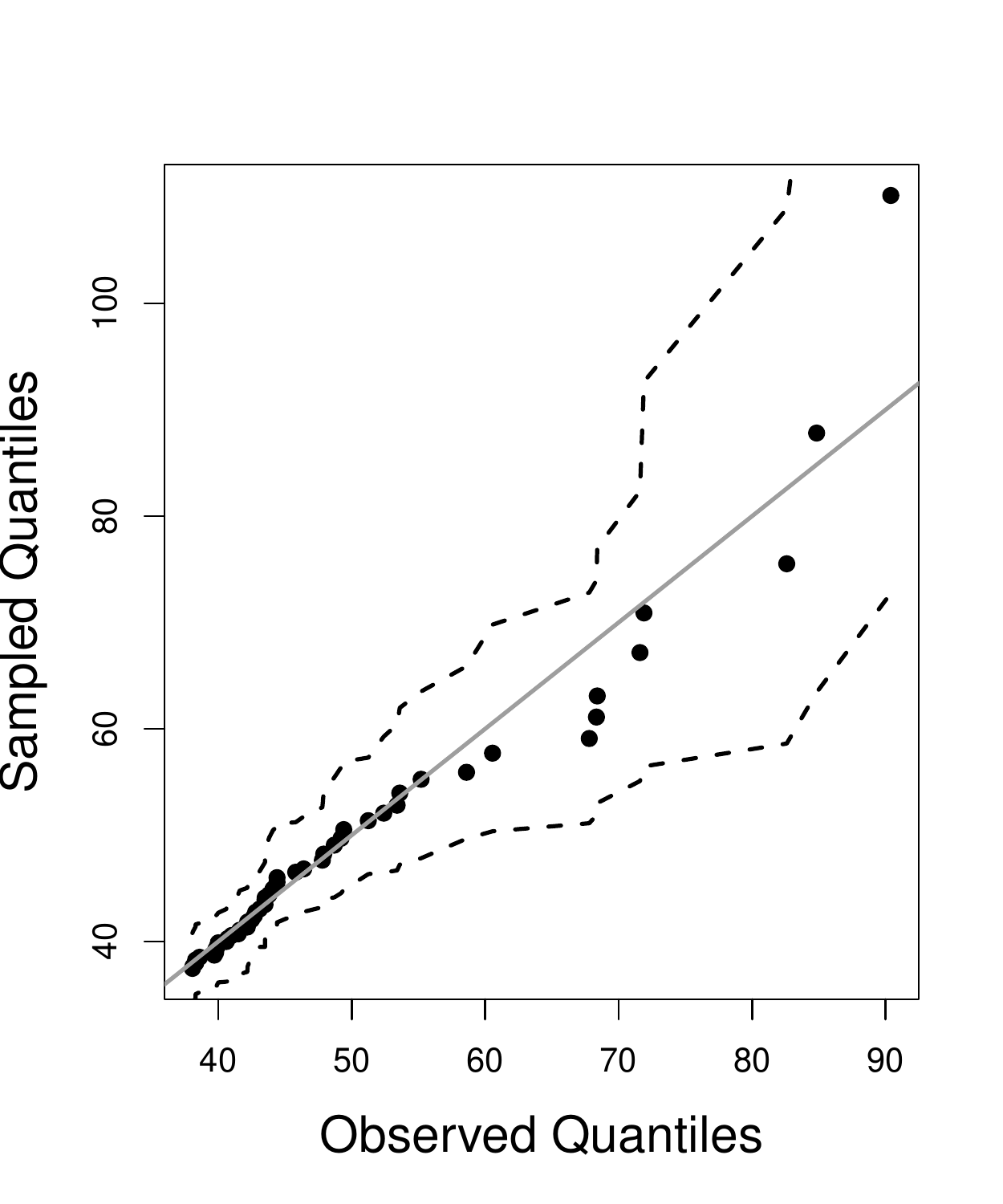}
\includegraphics[width=0.32\textwidth]{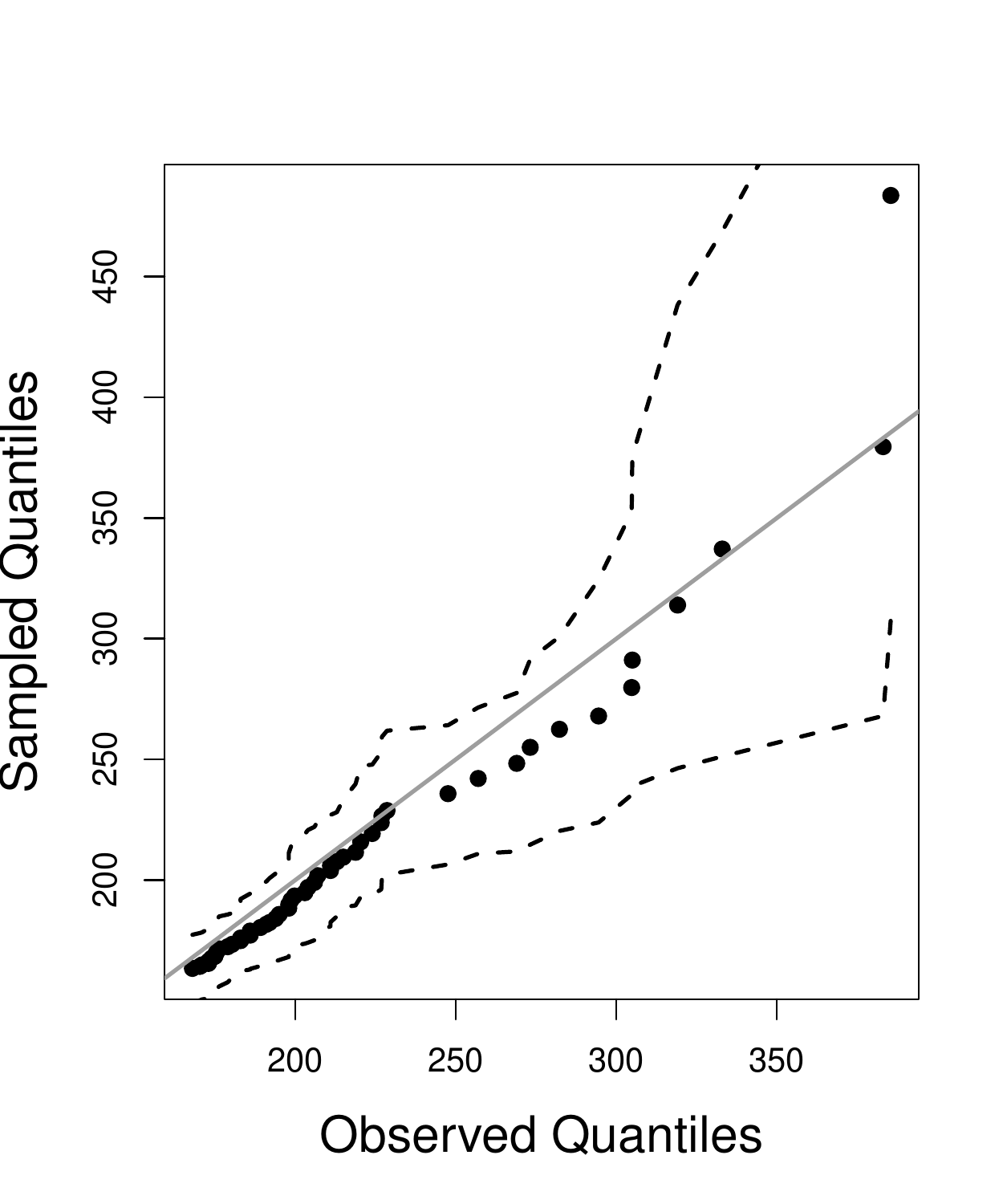}
\includegraphics[width=0.32\textwidth]{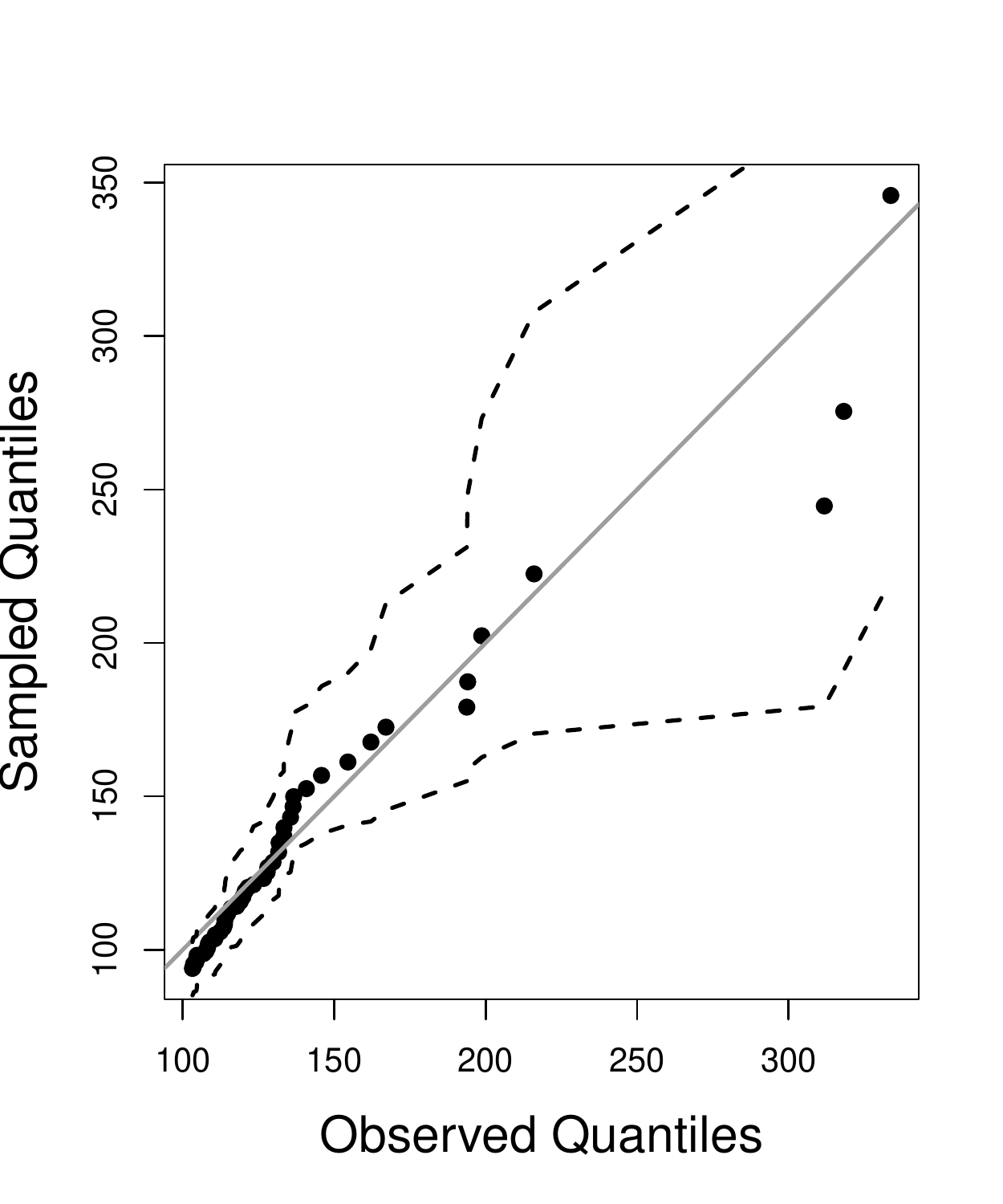}
\includegraphics[width=0.32\textwidth]{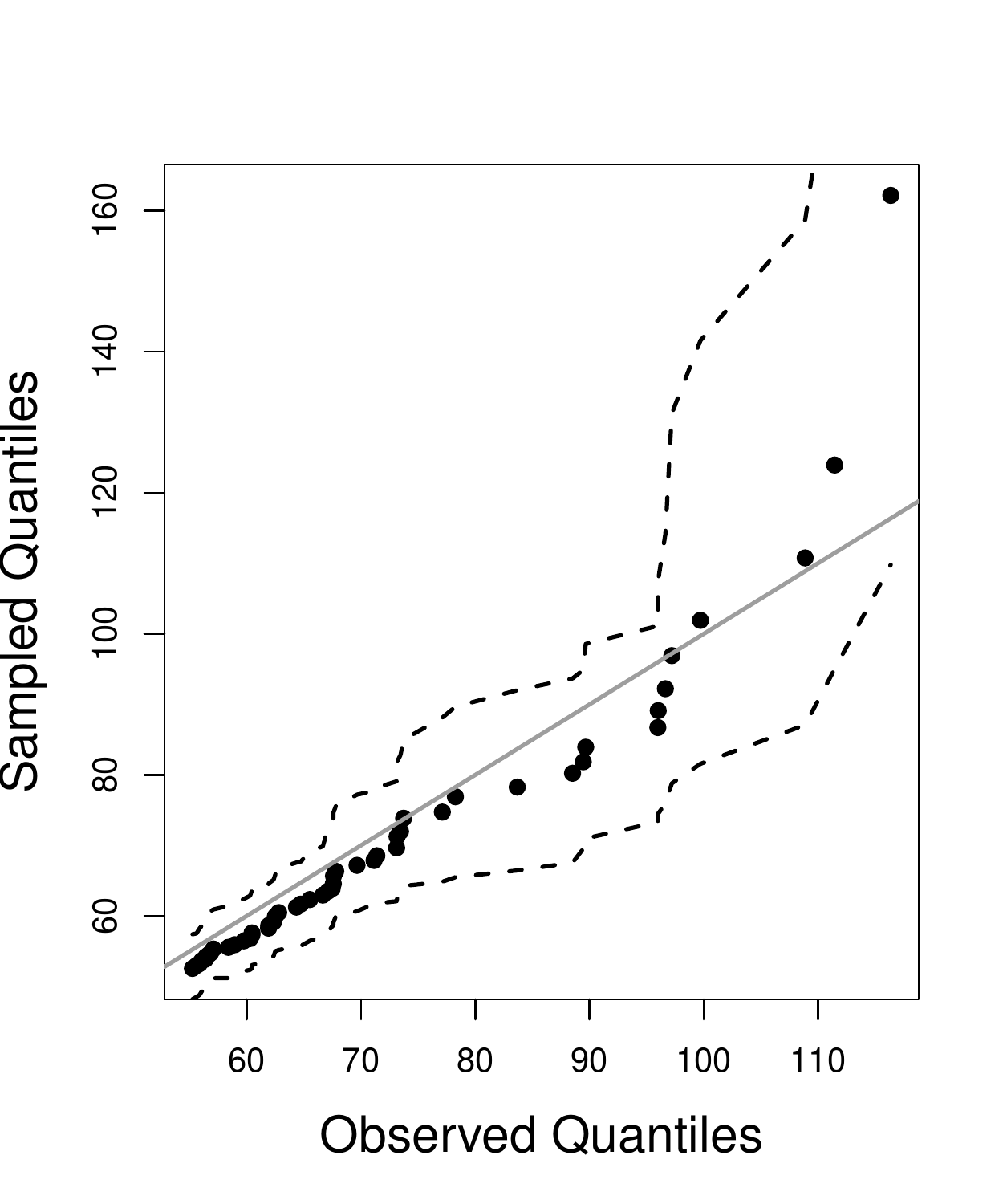}
\includegraphics[width=0.32\textwidth]{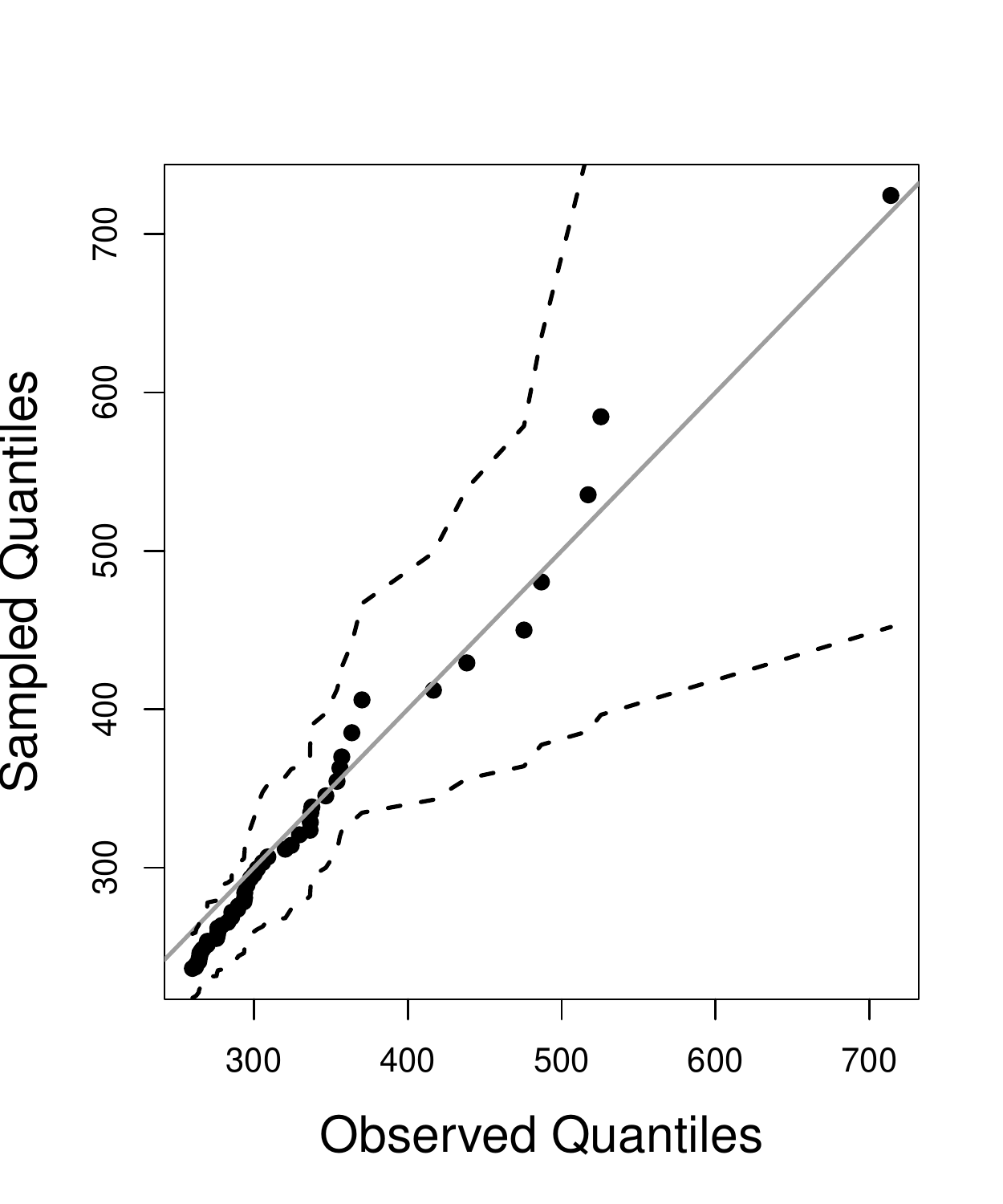}
\caption{Quantile-quantile plots for the fifty largest observations of $\max_{j\in\mathcal{G}} X_j$ (top) and $||\mathbf{X}_{\mathcal{G}}||_2$ (bottom) for the groups of gauges $\mathcal{G}_1 =\{3,4,16,20,24\}$ (left), $\mathcal{G}_2 =\{1,27,36,41,43\}$ (middle) and $\mathcal{G}_3 =\{6,15,21,25,33\}$ (right). The dashed lines correspond to the central 95\% sampling intervals.}
\label{fig:CheckDependence}
\end{figure}

The next step is to confirm that the extremal dependence structure of the simulated values agrees with that of the observed river flow levels. Figure~\ref{fig:ResultsPairs} indicates that our approach performs well at capturing pairwise extremal dependence. To examine this aspect in higher dimensions, we select three groups, each comprising five of the $K=45$ gauges in Figure~\ref{fig:Locations}. The groups are $\mathcal{G}_1 =\{3,4,16,20,24\}$, $\mathcal{G}_2 =\{1,27,36,41,43\}$ and $\mathcal{G}_3 =\{6,15,21,25,33\}$, and gauges within a group are similar regarding their maximum observed river flow levels. For each group $\mathcal{G}_j$~($j=1,2,3$), we consider two summary measures: (i) $\max_{k\in\mathcal{G}_j} X_k$, the maximum river flow across the gauges, and (ii)~$||\mathbf{X}_\mathcal{G}||_2= \left(\sum_{k\in\mathcal{G}_j} \mathbf{X}_k^2\right)^{1/2}$, the $L_2$-norm as a summary of aggregated river flow. As for the marginal distributions, we again focus on the fifty largest order statistics for $\max_{k\in\mathcal{G}_j} X_k$ and~$||\mathbf{X}_\mathcal{G}|$. Figure~\ref{fig:CheckDependence} shows that the observed order statistics for $\max_{k\in\mathcal{G}_j} X_k$ and~$||\mathbf{X}_\mathcal{G}||$ lie within the central 95\% sampling interval for all three groups. We conclude that our sampling algorithm in Section~\ref{sec:Generation} generates samples with similar characteristics to the extreme events of the UK river flow data. 

\begin{figure}[t!]
\centering
\includegraphics[width=0.48\textwidth, trim={0cm 2cm 0cm 2cm}]{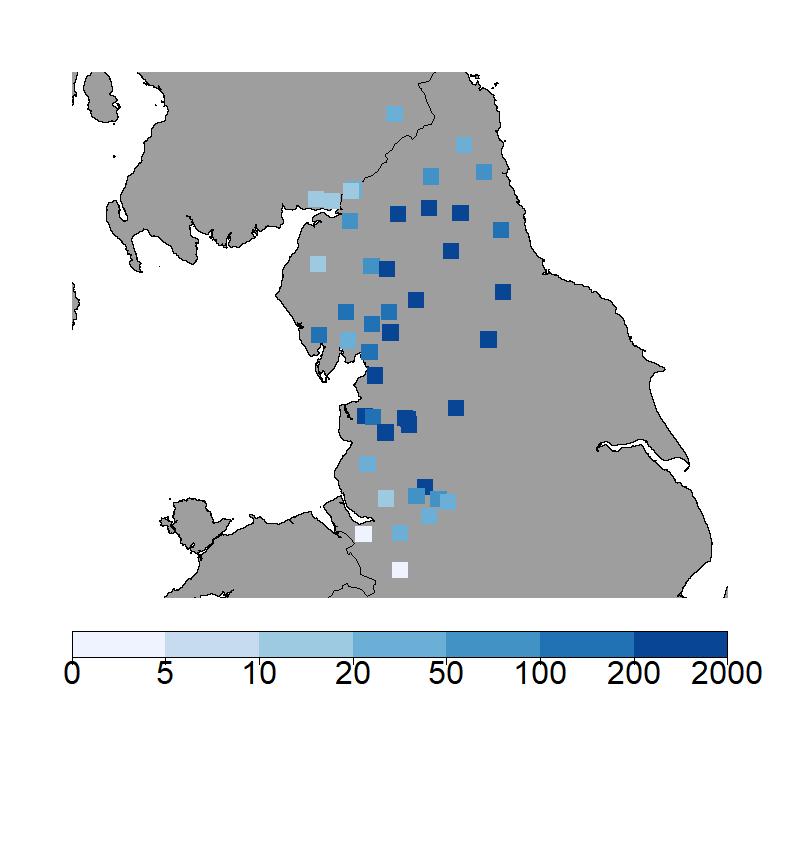}
\includegraphics[width=0.48\textwidth, trim={0cm 2cm 0cm 2cm}]{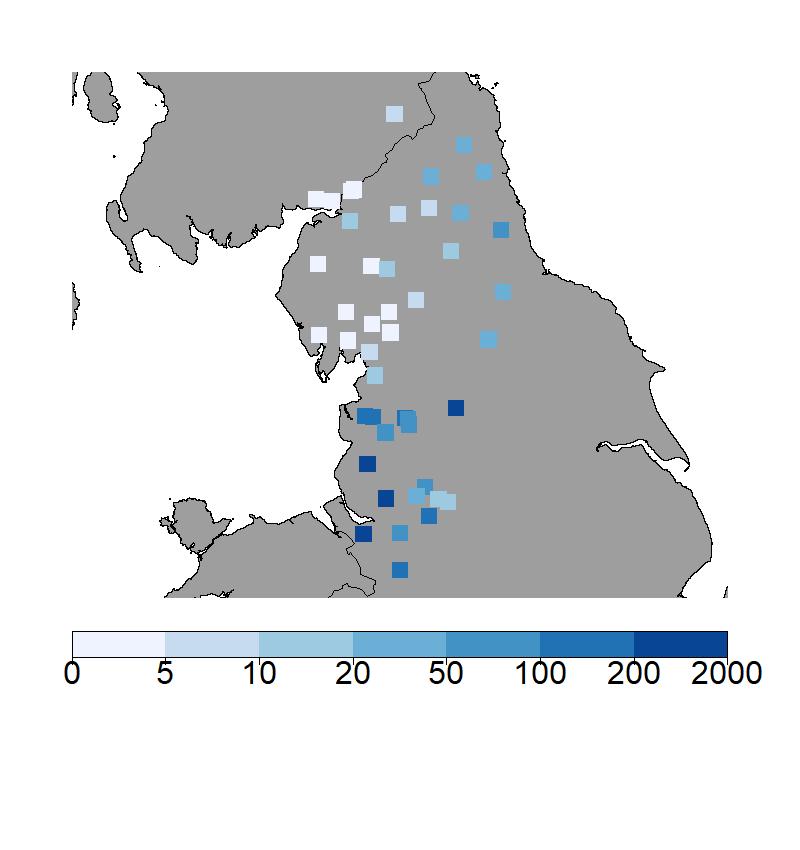}
\includegraphics[width=0.48\textwidth, trim={0cm 2cm 0cm 2cm}]{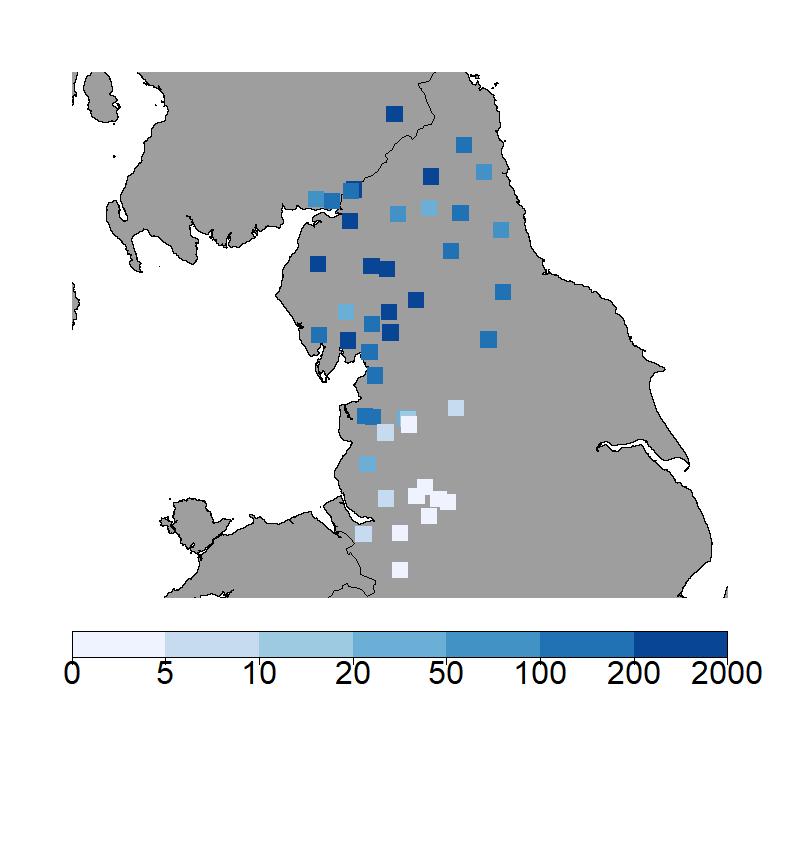}
\includegraphics[width=0.48\textwidth, trim={0cm 2cm 0cm 2cm}]{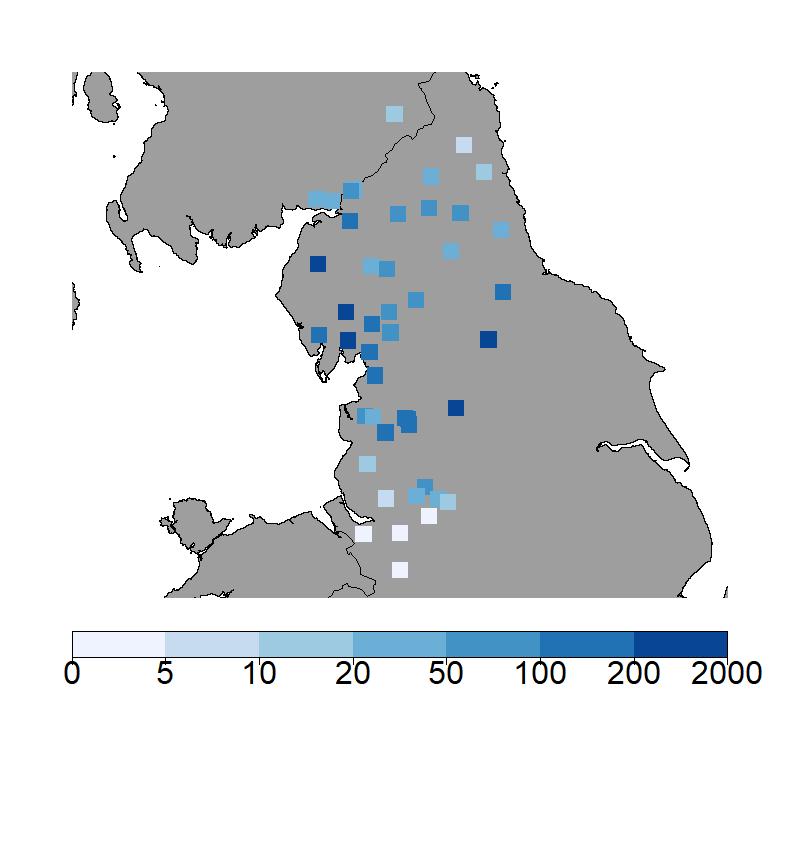}
\caption{Illustration of four simulated extreme events. The colouring shows the severity of the event at each gauge, in terms of it exceeding the estimated return levels. For instance, the darkest colour corresponds to the 200-year level being exceeded.}
\label{fig:ReturnLevelMaps}
\end{figure}

After verifying that our generative framework produces extreme events that exhibit properties similar to the observed extremes, we investigate the spatial structure of the simulated hazard events. We generate a set with 4,400 events, corresponding to a time window of 200 years. Figure~\ref{fig:ResultsPairs} illustrates the spatial pattern of the four most extreme simulated events in terms of $||\tilde{\mathbf{X}}||_2$. In the top left panel, the most extreme values are found in the centre of study region and almost all gauges observe severe river flow. The top right panel shows an event that predominately affects the southern half of the study region. Finally, the remaining two plots affect different partly different gauges in the northern half of the study region. As stated in Section~\ref{sec:Marginals}, we quantify severity using the concept of $\tau$-year event. The generated hazard event set includes 200-year events for 41 of the 45 gauges, and some of these are visible in Figure~\ref{fig:ResultsPairs}. The generated hazard event set can now be used as input for catastrophe models, for instance, to calculate insurance premiums.

\subsection{Comparison to existing methods}

\citet{Keef2013} and \citet{Quinn2019} generate hazard event sets using conditional extremes models. We briefly outline their framework before comparing its performance to our generative approach described in Section~\ref{sec:Generation}. The first step in \citet{Keef2013} is to transform the marginal distributions of $\mathbf{X}$ to common Laplace margins; this is similar to our approach, where we transform to Fr\'{e}chet margins. Let~$\mathbf{Y}$ denote the random vector $\mathbf{X}$ after the marginal transformation. \citet{Heffernan2004} show that under a relatively weak assumption, and for a sufficiently high threshold $v_k$, the distribution of $\mathbf{Y}_{-k} \mid (Y_k>v_k)$ can be approximated as
\begin{equation}
\mathbf{Y}_{-k} \mid (Y_k > v_k) = \bm\alpha_k Y_k + Y_k ^{\bm\beta_k} \mathbf{Z}_k\qquad ~(k=1,\ldots,K),
\label{eq:Keef2013}
\end{equation}
where $\mathbf{Y}_{-k}$ refers to $\mathbf{Y}$ without the $k$-th component, $\bm\alpha_k$ and $\bm\beta_k$ are $K-1$-dimensional parameter vectors, which vary with the conditioning variable $Y_k$, and $\mathbf{Z}_k$ is a $K-1$-dimensional random vector that is independent of $Y_k$. 

Given estimates $(\hat{\bm\alpha}_1,\hat{\bm\beta}_1), \ldots, (\hat{\bm\alpha}_K,\hat{\bm\beta}_K)$ and fitted distributions for $\mathbf{Z}_1,\ldots,\mathbf{Z}_K$, synthetic observations for $\mathbf{Y}$ are generated as follows:
\begin{enumerate}
\item Sample the index $k$ of the conditioning site and a value $y_k$ for $Y_k$ with $y_k>v_k$.
\item Draw a realisation $\mathbf{z}^*_k$ from the fitted distribution of $\mathbf{Z}_k$.
\item Derive the simulated values for the other $K-1$ sites with
$\mathbf{y}_{-k} = \hat{\bm\alpha}_k y_j + y_j ^{\hat{\bm\beta}_k} \mathbf{z}^*_k$.
\end{enumerate}
There a few more intricacies to this approach and we refer the reader to \citet{Keef2013} for details. The R package \texttt{texmex} was used for most of the operations, including the estimation of the marginal GPD parameters $(\sigma_k,\xi_k)$ in \eqref{eq:GPD} and the parameter vectors $\bm\alpha_k$ and $\bm\beta_k$~($k=1,\ldots,K$). The vector $\mathbf{z}_k^*$ in Step 2 of the algorithm was sampled from the empirical distribution function of~$\mathbf{Z}_k$. We set the threshold $v_k$ in \eqref{eq:Keef2013} to the empirical 93\% quantile of $Y_k$~($k=1,\ldots,K$); we explored alternatives and they gave results similar to those presented in the following.

\begin{figure}
\centering
\includegraphics[width=0.32\textwidth]{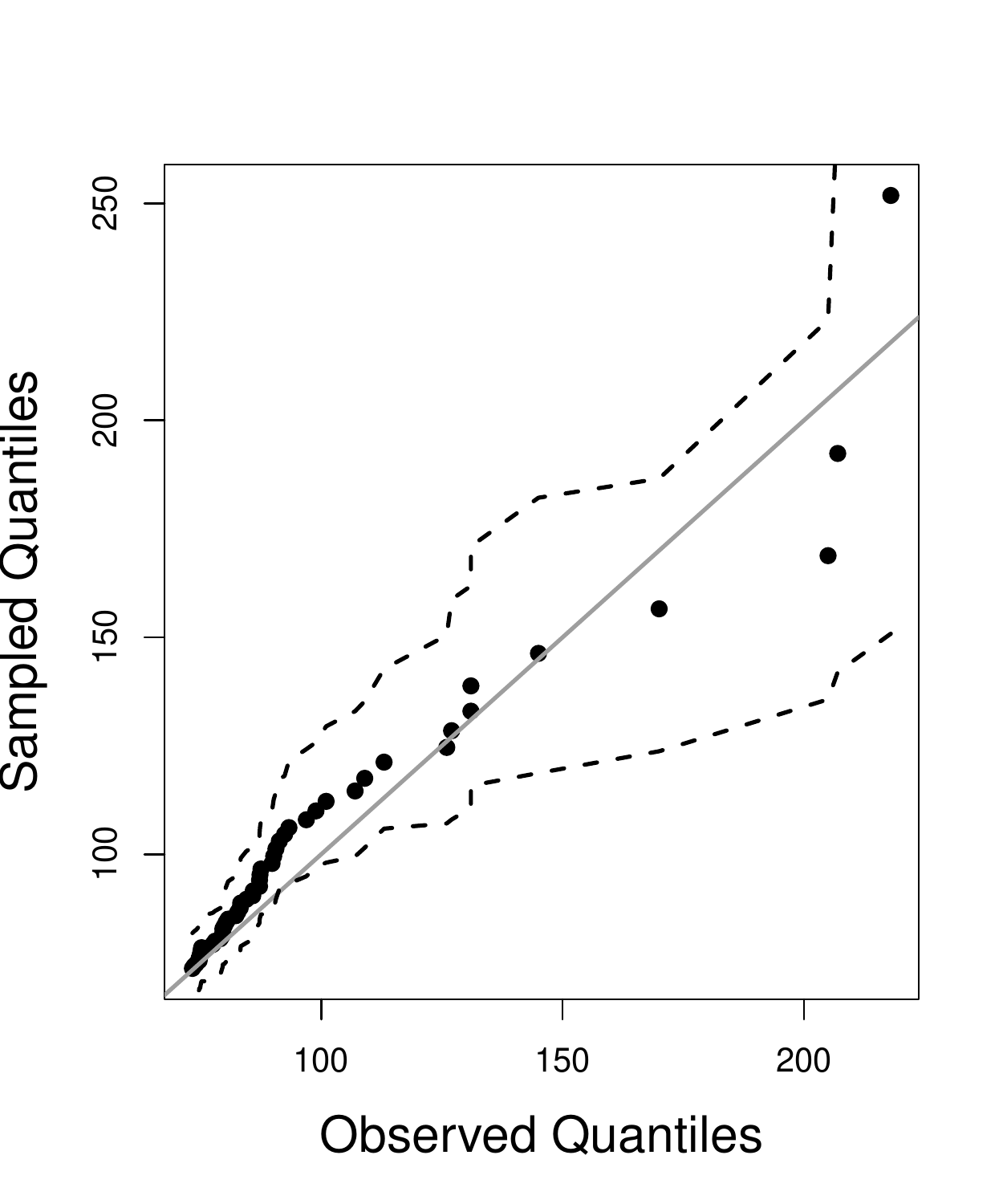}
\includegraphics[width=0.32\textwidth]{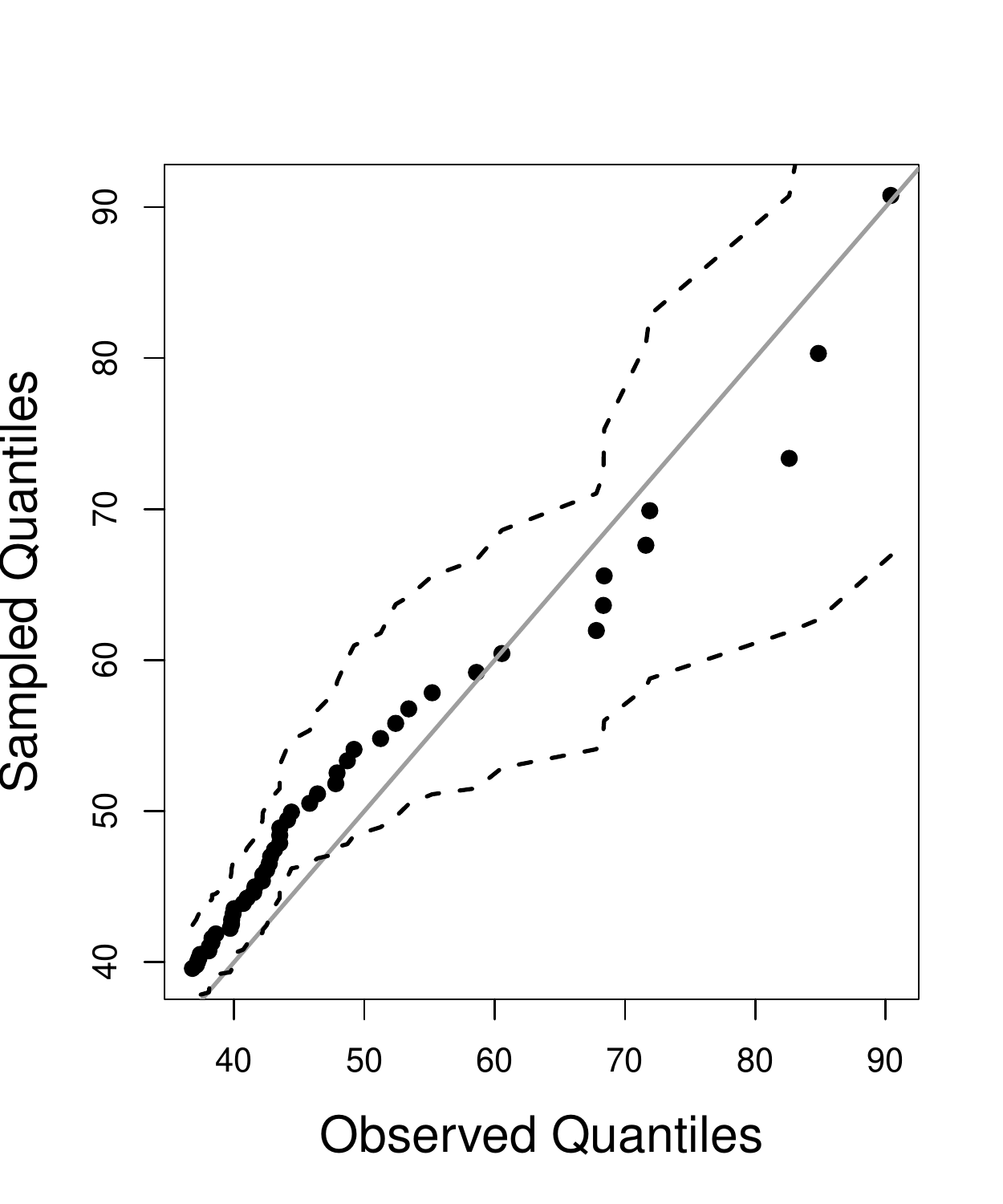}
\includegraphics[width=0.32\textwidth]{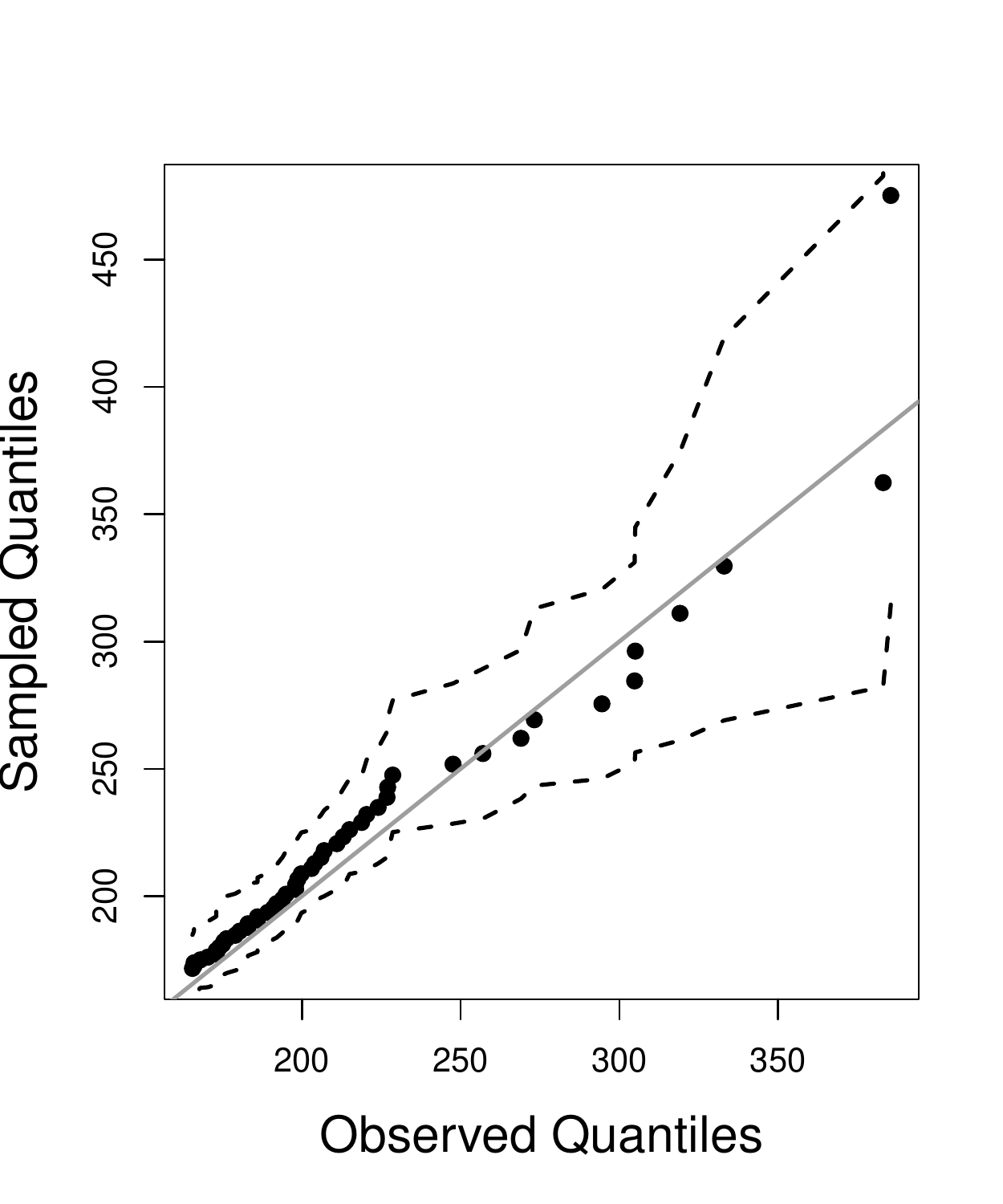}
\caption{Quantile-quantile plots for the fifty largest observations of $\max_{j\in\mathcal{G}} X_j$ for the groups of gauges $\mathcal{G}_1 =\{3,4,16,20,24\}$ (left), $\mathcal{G}_2 =\{1,27,36,41,43\}$ (middle) and $\mathcal{G}_3 =\{6,15,21,25,33\}$ (right). The simulated observations are obtained using the sampling algorithm by \citet{Keef2013}. The dashed lines correspond to the central 95\% sampling intervals.}
\label{fig:Comparison}
\end{figure}

As in Section~\ref{sec:Validation}, 100 hazard event sets are simulated for the random vector $\mathbf{X}$, and performance is assessed based on the summaries $\max_{k\in\mathcal{G}_j} X_k$ and~$||\mathbf{X}_\mathcal{G}||_2$ for the three groups of gauges. Figure~\ref{fig:Comparison} shows the results for $\max_{k\in\mathcal{G}_j} X_k$, and the plots for $||\mathbf{X}_\mathcal{G}||_2$ are provided in the supplementary material. While the plots show a reasonable agreement between the simulated and observed values of $\max_{k\in\mathcal{G}_j} X_k$, they also indicate a bias for the non-extreme but high values of $\max_{k\in\mathcal{G}_j} X_k$~($j=1,2,3$). The same result is found for $||\mathbf{X}_\mathcal{G}||_2$, and the bias remains when we increase $v_k$ to the 96\%~empirical quantile of~$Y_k$.

\begin{figure}
\centering
\includegraphics[width=0.49\textwidth, trim={0cm 2cm 0cm 2cm}]{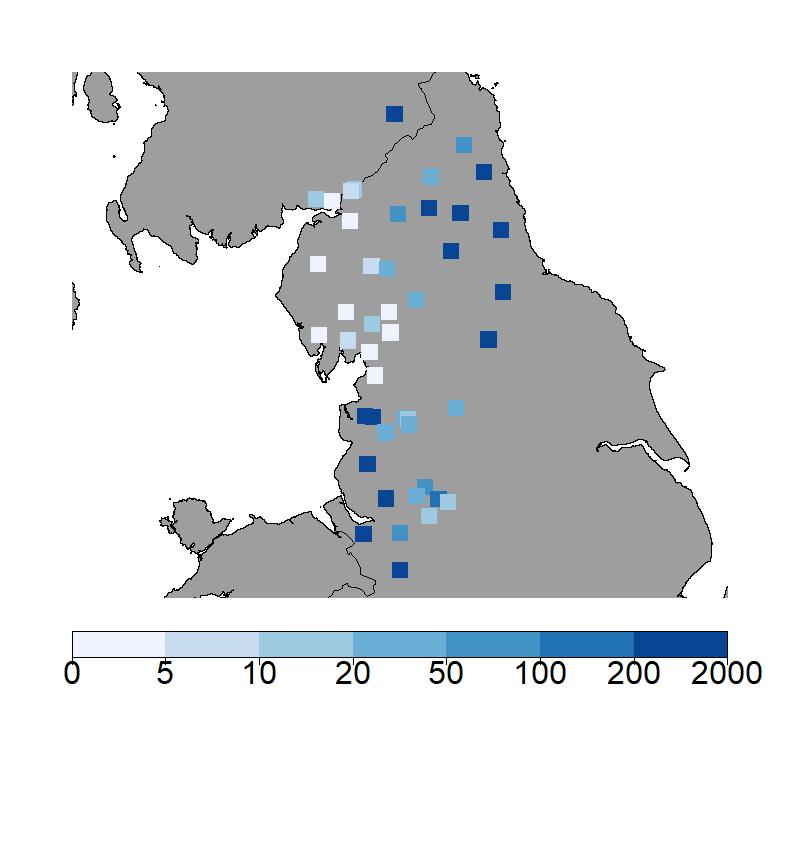}
\includegraphics[width=0.49\textwidth, trim={0cm 2cm 0cm 2cm}]{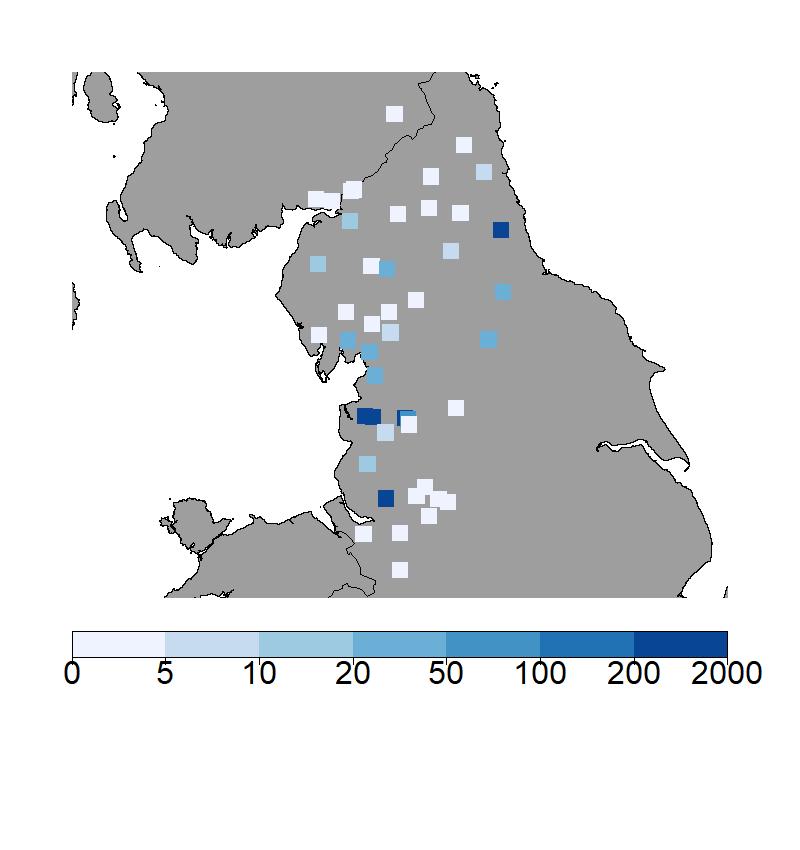}
\caption{Illustration of the two most severe extreme events generated by the approach of \citet{Keef2013}. The colouring shows the severity of the event at each gauge, in terms of it exceeding the estimated return levels. For instance, the darkest colour corresponds to the 200-year level being exceeded.}
\label{fig:Comparison2}
\end{figure}

We again conclude our analysis by investigating the spatial structure of the generated extreme events. Figure~\ref{fig:Comparison2} shows two of the four most extreme events generated by the conditional extremes approach. While the extreme event in the left panel appears realistic, and has similarities to one of the events in Figur~\ref{fig:ReturnLevelMaps}, the synthetic extreme event in the right panel is highly unlikely -- the event corresponds to very extreme river flow across spatially distant gauges, with the gauges between them experiencing much less extreme river flow. Consequently, the approach by \citet{Keef2013} exhibits limitations in the context of our UK river flow application.

\section{Discussion}
\label{sec:Discussion}

This paper proposed a novel generative framework, based on a principal component analysis of the most severe events, to obtain hazard event sets of extreme river flow across 45 gauges in northern England and southern Scotland. We used the approach, common to extremes, of separating the modeling of marginal distributions from the characterization of dependence. For marginal estimation, we used the clustering method by \citet{RohrbeckTawn2020} to reduce estimation variability; \cite{Huser2022} provides a summary of other spatial methods for estimating marginal extreme value models. Extremal dependence was analysed by the principal component decomposition of \citet{Cooley2019}. We find a close link between the first eigenvectors of the TPDM and known geographical / climatological features of the study region. The robustness of this analysis was assessed using a non-parametric bootstrap --- the combined interpretation of the first six eigenvectors is usually the same, with potential permutations of the fourth to sixth eigenvector. 

Our generative framework is based on a statistical model for the extremal principal components. We reduce the dimension by modelling only the leading extremal principal components using a kernel density estimate, and use an empirical estimate for the remaining components. Cross-validation is used to select the number of principal components to model, and bootstrapping is used to account for uncertainty. We found good agreement between the generated and observed extreme events. Because the methods required to generate samples are computationally efficient, the method can be easily implemented by practitioners in insurance or engineering.

We assumed that the data are stationary, both in their marginal distributions and dependence structure. Non-stationarity in the marginal distributions could be modelled using the approaches by \cite{Davison1990}, \cite{Eastoe2009}, or \cite{Eastoe2019}, amongst others. Extrapolating non-stationarity into the future requires assumptions about future behavior, and accounting for non-stationarity in the dependence structure is much more challenging.

While extreme flood events are spatial by nature, the application herein is rather non-standard since we consider a river network with disparate catchments. As such, the dependence structure is highly complex, and any spatial extremes model has to incorporate both geographical and hydrological distances \citep{Asadi2015}. We attempted to use the R package {\tt mvpot} to compare our method to a Brown-Resnick process with extremal dependence characterized by an anisotropic covariance function, and distance based on latitude and longitude coordinates. We found the estimated parameters were highly sensitive to the initial parameters set for the optimization problem, and the produced samples seemed unrealistic.

Our generative framework can be applied to other study regions, seasons and environmental/hydrological variables. The number of sites that can be handled efficiently by our approach depends on the underlying dependence structure. We found that $m=7$ performed best in our application, but a larger number of components may be required in other settings. If a large number of principal components is required, the kernel-density estimate may be replaced by a mixture of von Mises-Fisher distributions, with the number of mixture components being estimated using clustering.

Finally, developments in machine learning are providing alternative methods for modeling dependence in high-dimensional settings. Recently, Generalized Adversarial Networks (GANs) have been tailored for estimating extremal dependence \citep{Bhatia2020, Boulaguiem2022}. While such approaches seem to be quite powerful, they are difficult to interpret. We believe the ability to relate the leading vectors of the eigenbasis to known geographic features as a strong advantage of our approach.

\subsection*{Acknowledgement}
Parts of this research were conducted while Christian Rohrbeck was beneficiary of an AXA Research Fund postdoctoral fellowship grant. Dan Cooley's work was partially supported by the US National Science Foundation grant DMS-1811657. The daily river flow data were obtained through the National River Flow Archive. We thank the editor, associate editor and two anonymous referees for helpful comments and suggestions that have helped us to improve this work.

\singlespacing

\bibliographystyle{apalike}
\bibliography{References}

\end{document}